%% file: main.tex
\documentclass[ALICE,manyauthors]{cernphprep}
\usepackage[comma,square,numbers,sort&compress]{natbib}
\usepackage{hyperref}
\usepackage{lineno}
\usepackage{xspace}
\usepackage[T1]{fontenc}
\usepackage {orcidlink}
\begin{document}
\input{commands.tex}

\begin{titlepage}
\PHyear{2026}       
\PHnumber{006}      
\PHdate{13 January}  

\title{Measurements of the production of W$^{\pm}$ and Z$^0$ bosons in pp collisions at $\sqrt{s} = 13$ TeV}
\ShortTitle{Production of W$^{\pm}$ and Z$^0$ bosons in pp collisions at \s = 13 \TeV}   

\Collaboration{ALICE Collaboration\thanks{See Appendix~\ref{app:collab} for the list of collaboration members}}
\ShortAuthor{ALICE Collaboration} 

\begin{abstract}
Measurements of the production of the W$^{\pm}$ and Z$^0$ bosons at midrapidity in pp collisions at $\sqrt{s} = 13$ TeV with ALICE at the Large Hadron Collider (LHC) are presented. The W$^{\pm}$ and Z$^0$ bosons are detected via their (di)electronic decay channels, with the electron reconstruction performed in the midrapidity region ($|y|<$ 0.6). The $p_{\textrm{T}}$-integrated and $p_{\rm T}$-differential production cross sections of electrons from W$^{\pm}$ decays in the interval 30 $<p_{\textrm{T}}<$ 60 GeV/$c$, as well as the $p_{\textrm{T}}$-integrated production cross section of Z$^0$ bosons, are measured. The results are described by perturbative QCD calculations using different sets of parton distribution functions. The production of W$^{\pm}$ bosons and azimuthally correlated associated hadrons is also measured as a function of the charged-particle multiplicity for the first time at the LHC. The former increases approximately linearly with the charged-particle multiplicity, while for the latter, there are hints of a faster-than-linear increase. These observations are compared with theoretical calculations. 
\end{abstract}
\end{titlepage}

\setcounter{page}{2} 


\section{Introduction} 

The quark--gluon plasma (QGP)~\cite{Bazavov:2011nk} is a state of strongly--interacting QCD matter present under extreme temperature and energy density conditions reached in high-energy heavy-ion collisions.
In such collisions, a strong suppression of the particle production at high transverse momentum ($p_{\rm T}$), as well as a significant azimuthal anisotropy of low-$p_{\textrm{T}}$ particles, are considered up to now as signatures of interactions of the partons with the created hot and dense medium.
However, recent measurements of two-particle angular correlations in high-multiplicity pp, p--A and d--A collisions revealed a non-zero $v_{2}$ for not only light-flavour hadrons~\cite{ATLAS:2015hzw, ALICE:2012eyl, ATLAS:2016yzd, CMS:2013jlh, ALICE:2013snk, CMS:2014und, PHENIX:2013ktj, STAR:2015kak, PHENIX:2015idk} but also heavy-flavour hadrons~\cite{ALICE:2018gyx, CMS:2018loe, CMS:2018duw, ALICE:2017smo, CMS:2020qul}, which can be described by models that incorporate a hydrodynamical evolution of the medium. These results can also be reproduced by models that, without invoking final-state effects, include multiparton interactions (MPI) together with colour reconnection (CR)~\cite{Sjostrand:2014zea, OrtizVelasquez:2013ofg}, overlapping strings~\cite{Bierlich:2014xba}, or string percolation~\cite{Bautista:2015kwa}. Moreover, no clear evidence of a strong suppression of the high-$p_{\textrm{T}}$ particle production has been observed so far in small collision systems (pp, p--A and d--A collisions). The PHENIX collaboration recently reported a 20\% suppression of high-$p_\textrm{T}$ $\pi^{0}$ yields in d--Au collisions with high activity using the measurement of the nuclear modification factor normalised to that for direct photons to avoid event selection bias. These results were interpreted as a possible hint of final-state effects~\cite{PHENIX:2023dxl}.
The hadron production including soft-$p_\textrm{T}$ ranges has also been studied as a function of the charged-particle multiplicity. The ALICE measurements of the self-normalised yields as a function of charged-particle multiplicity at midrapidity in high-multiplicity pp collisions at $\sqrt{s}=$ 5.02, 7 and 13 {\TeV} show a faster-than-linear increase for both light-flavour~\cite{ALICE:2019dfi, ALICE:2019avo} and heavy-flavour hadrons~\cite{ALICE:2015ikl, ALICE:2020msa, ALICE:2023xiu}, which is found more pronounced for high-$p_{\textrm{T}}$ particles~\cite{ALICE:2020msa}.
Models incorporating multiparton interactions and colour-reconnection effects describe qualitatively this behaviour, as well as the observed enhancement in the baryon to meson ratios in high-multiplicity pp collisions~\cite{ALICE:2021npz}. At LHC energies, roughly between four to ten partonic interactions occur per event~\cite{Sjostrand:2004pf}. 
In the CR mechanism, partons from different partonic interactions can reconnect through their colour charge and combine into hadrons~\cite{Lonnblad:2023stc,Gustafson:2009qz}.
On the other hand, a linear increase is observed in the measurement of J/$\psi$ mesons at forward rapidity, where a pseudorapidity gap between the J/$\psi$ mesons and the midrapidity charged-particle multiplicity estimator is present~\cite{ALICE:2021zkd}. Therefore, it remains unclear whether the observed increase in particle production at high multiplicity in small collision systems is a genuine MPI effect or driven by correlations between the measured hadron and the charged-particle multiplicity (autocorrelations). Measurements of electroweak bosons and high-$p_{\rm T}$ particles produced back-to-back are particularly interesting, since they allow to compare the effects on very different probes, one subject to the strong interaction and the other to the electromagnetic one.
 
Electroweak bosons, W$^{\pm}$ and Z$^{0}$/$\gamma^{*}$ (referred to as Z$^{0}$ in the remainder of the publication), are predominantly generated through the Drell-Yan process in which quarks and antiquarks annihilate in the early stage of the collisions~\cite{Schott:2014sea}.
Since they are weakly-interacting particles insensitive to the strong interaction, their production is expected to be less affected by final-state effects and CR effects than the hadron production. Measurements of the multiplicity dependence of W$^{\pm}$- and Z$^{0}$- boson production can, therefore, provide important insights into the role of MPI and CR for hadron production, and serve as a reference probe for assessing the QGP formation in the small colliding systems. Due to their large masses ($M_{\rm W^\pm}$= 80.3692 $\pm$ 0.0133 GeV/$c^{2}$ and $M_{\rm Z^0}=$ 91.1880 $\pm$ 0.0020 GeV/$c^{2}$)~\cite{ParticleDataGroup:2024cfk}, their production cross section can be described by perturbative QCD (pQCD) in high-energy pp collisions~\cite{Anastasiou:2003ds, Melnikov:2006kv}.
In theoretical calculations, the uncertainties arise from the strong-interaction coupling ($\alpha_{s}$), the choice of QCD scales for renormalisation ($\mu_{\textrm{R}}$) and factorisation ($\mu_{\textrm{F}}$), and the momentum distributions of partons in the proton (PDFs). The PDFs, determined through a phenomenological analysis based on experimental data~\cite{Alekhin:2017kpj,Hou:2019efy,Harland-Lang:2014zoa, NNPDF:2017mvq}, are a particularly crucial ingredient of the production cross section calculations. 
Therefore, measuring the production cross sections for W$^{\pm}$ and Z$^{0}$ bosons is important to test QCD and constrain the PDFs. The production of W$^{+}$ and W$^{-}$ is in particular sensitive to the light quark PDFs~\cite{Paukkunen:2010qg} since  they are mainly produced by interactions between d$\bar{\textrm{u}} \rightarrow$ W$^{-}$ and u$\bar{\textrm{d}} \rightarrow$ W$^{+}$ processes, and their asymmetry is sensitive to the down-to-up quark ratio in the proton. At the LHC, the production of W$^{\pm}$ and Z$^{0}$ bosons, integrated over the charged-particle multiplicity, has been extensively measured in pp collisions by the ATLAS, CMS and LHCb experiments, see~\cite{ATLAS:2019fyu, ATLAS:2018pyl, LHCb:2023qav, LHCb:2025msn, CMS:2011aa, LHCb:2014liz, LHCb:2015okr, ATLAS:2016nqi, LHCb:2015mad, CMS:2016mwa, ATLAS:2015iiu, ATLAS:2016fij, CMS:2020cph, LHCb:2021huf} and references therein.
In this article, the first ALICE measurements of W$^\pm$- and Z$^0$-boson production via the (di)electronic decay channel ($\textrm{W}^{+} \rightarrow \textrm{e}^{+} \nu$,
$\textrm{W}^{-} \rightarrow \textrm{e}^{-} \overline{\nu}$,
$\textrm{Z}^0 \rightarrow \textrm{e}^{+}\textrm{e}^{-}$) at midrapidity in pp collisions at $\sqrt{s}=$ 13 TeV are reported.
The $Z^0$ bosons are reconstructed through the invariant mass of $\textrm{e}^{+}\textrm{e}^{-}$ pairs and the production cross section is measured in the fiducial kinematic region 60 $< m_{{\rm ee}}< $ 108 GeV/$c^{2}$, where one electron is in the range 30 $< p_{\textrm{T}}< $ 60 GeV/$c$ and $|y|<$ 0.6. 
The $p_{\textrm{T}}$-differential and $p_{\textrm{T}}$-integrated production cross sections for electrons (positrons) from W$^{-}$ (W$^{+}$) are measured in the range 30 $< p_{\textrm{T}}<$ 60 GeV/$c$ for $|y|<$ 0.6. The measured production  cross sections for $\textrm{e}^{\pm} \leftarrow \textrm{W}^{\pm}$ and Z$^0$ decaying
into dielectrons are compared with theoretical calculations based on NLO pQCD using recent PDFs~\cite{Yan:2022pzl,NNPDF:2021njg}. For the first time at the LHC, the production of $\textrm{e}^{\pm} \leftarrow \textrm{W}^{\pm}$ and associated hadrons, extracted from electron-hadron azimuthal correlations, is studied as a function of the charged-particle multiplicity and compared with PYTHIA 8 calculations.

\section{Experimental apparatus and data--taking conditions}

The analysis is performed using pp collisions at $\sqrt{s}=$ 13 TeV recorded by ALICE from 2016 to 2018. A detailed description of the apparatus and its performance can be found in Refs.~\cite{Aamodt:2008zz, Abelev:2014ffa}. The ALICE detector consists of a central barrel at $|\eta|<$ 0.9, a forward muon spectrometer ($-4 < \eta < -2.5$), and a set of detectors for triggering and event characterisation at forward and backward rapidities. 
The central barrel detectors are positioned inside a large solenoid magnet, providing a uniform magnetic field of 0.5\,T.

The main subdetectors used in the analysis are the Inner Tracking System (ITS)~\cite{Aamodt:2010aa}, the Time Projection Chamber (TPC)~\cite{Alme:2010ke} and the Electromagnetic Calorimeter (EMCal)~\cite{ALICE:2022qhn} located in the central barrel. The ITS is the innermost detector, located at radii between 3.9 and 43 cm from the nominal interaction point. It consists of three subdetectors, the Silicon Pixel Detector (SPD), the Silicon Drift Detector (SDD), and the Silicon Strip Detector (SSD). Each detector has two cylindrical silicon layers in full azimuthal coverage surrounding the beam pipe. The ITS is used for charged-particle track reconstruction. It contributes to the reconstruction of the collision vertex (primary vertex) and the secondary vertices of weakly decaying particles, and  allows  precise measurements of the distance of closest approach of the tracks to the primary vertex.

Charged particles produced in the collisions are mainly reconstructed by the TPC with up to 159 three-dimensional space points per track. The TPC covers the full azimuthal angle $\varphi$ in the pseudorapidity range $|\eta|<$0.9. It enables the identification of charged particles via the specific ionisation energy loss (d$E$/d$x$). Electrons are identified exploiting the information of the EMCal. The EMCal is a layered lead-scintillator sampling electromagnetic calorimeter that covers an acceptance  of $\vert \eta \vert < 0.7$ in pseudorapidity and $\Delta \varphi = 107^\circ$ in azimuth. 
The smallest segmentation of the EMCal is a tower, which has a dimension of 0.0143 $\times$ 0.0143 rad$^{2}$ in the $\eta \times \varphi$ direction. The energy resolution of the EMCal is $\sigma_{E}/E=$ 2.9\%/$E\oplus$9.5\%/$\sqrt{E}\oplus$1.4\%, where $E$ is the energy in GeV~\cite{ALICE:2022qhn}. 
In addition, the V0 detector which consists of two arrays of 32 scintillator tiles each, covering the pseudorapidity intervals 2.8 $<\eta<$5.1 (V0A) and $-$3.7 $<\eta<-$1.7 (V0C), serves as trigger detector and to reject offline beam-induced background using the V0 timing information~\cite{ALICE:2013axi}.

The analysis is based on a sample of events selected by a minimum bias (MB) trigger, which requires a coincident signal in the V0A and V0C. The contribution of events containing more than one collision (pileup events) is reduced by requiring a single primary vertex reconstructed by the ITS and the TPC within $\pm$10 cm from the nominal interaction point~\cite{Abelev:2014ffa}. 
In addition to the MB trigger, an EMCal trigger is required in order to enhance the number of highly energetic electrons in the sample.
The EMCal trigger is based on the sum of energy in a sliding window of 4 $\times$ 4 towers, with a threshold set at 9 GeV (EG1 trigger). The number of analysed EMCal triggered events after the event selection is 60 M, corresponding to an integrated luminosity of 7.8 pb$^{-1}$.
The integrated luminosity was obtained by normalizing the triggered events to the equivalent number of MB events using a trigger rejection factor determined with a data-driven method, based on the ratio of EMCal cluster energy distributions measured in MB and triggered events. The luminosity calibration of the MB trigger is based on the V0 reference cross section measured via van der Meer scans~\cite{ALICE-PUBLIC-2021-005}.

The charged-particle pseudorapidity density ($\textrm{d}N_{\textrm{ch}}/\textrm{d}\eta$) is measured in the pseudorapidity interval $|\eta|< 1$ using the number of tracklets in the SPD ($N_{\textrm{tracklets}}$)~\cite{ALICE:2015qqj}.
The tracklets are defined as combinations of two hits in the SPD layers, pointing to the primary vertex. The number of measured  tracklets depends on the collision vertex position due to the limited acceptance of the SPD, and on the number and location of the active SPD channels during the various data-taking periods.
These detector effects were corrected event-by-event using a data-driven approach~\cite{ALICE:2012pet, ALICE:2020msa}. 
Based on this method, the measured average number of tracklets is equalised to a reference value chosen as the maximum average tracklet number observed as a function of both collision vertex position and time. The conversion from $N_{\textrm{tracklet}}$ to $N_{\textrm{ch}}$ is estimated using a Monte Carlo simulation.
Primary charged particles are generated with the  PYTHIA 8.2 event generator~\cite{Sjostrand:2014zea}, and propagated through the detector material with the GEANT3 transport code~\cite{Brun:1082634}, using the same detector conditions over time as in the data.
The charged-particle pseudorapidity density is normalised to its average value ($\langle \textrm{d}N_{\textrm{ch}}/\textrm{d}\eta \rangle$ = 13.9) in INEL $> 0$ events~\cite{ALICE:2020swj}, defined as events with at least one charged particle in $|\eta|<$ 1. In the following, the normalised charged-particle multiplicity is expressed as $\textrm{d}N_{\textrm{ch}}/d\eta/\langle \textrm{d}N_{\textrm{ch}}/\textrm{d}\eta \rangle_{\rm INEL > 0}$.

\section{W$^\pm$- and Z$^0$-boson reconstruction}
The W$^{\pm}$ and Z$^{0}$ bosons are measured via their (di)electronic decay channels. The Z$^0$ bosons are reconstructed through the  invariant mass of the two decay electrons e$^+$e$^-$, while the signal extraction for W$^{\pm}$ bosons is performed from the single electron $p_{\textrm{T}}$ distribution. The tracks are required to have at least three hits in the ITS (out of 6), with at least one hit in either of the two SPD layers. Moreover, tracks with a distance of closest approach to the primary vertex of less than 3.2 cm along the beam axis and 2.4 cm in the transverse plane are selected. Additionally, tracks with at least 80 clusters reconstructed in the TPC (out of a total of 159) are chosen to ensure the selection of high-quality tracks.
Tracks passing through the fiducial volume of the EMCal are selected by requiring the rapidity of the electron to be $\vert y \vert < 0.6$ to avoid the edge of the detector.
The tracks are extrapolated from the TPC to the EMCal by taking into account the energy loss of particles in the detector material. EMCal clusters are then associated to the tracks based on the spatial matching between them and the extrapolated tracks~\cite{ALICE:2017nce}.
Electron candidates are identified using their ionisation energy loss (${\rm d}E/{\rm d}x$) in the TPC gas. A large fraction of hadron background is suppressed by requiring the measured ${\rm d}E/{\rm d}x$ to fullfill the condition $-1 < n_{\sigma}^{\textrm{TPC}} < 3$, where $n_{\sigma}^{\textrm{TPC}}$ is defined as its deviation from the expected value for electrons normalised by the detector resolution. Additional hadron rejection is achieved with selection criteria based on the ratio of the energy deposited by a charged track in the EMCal and its momentum measured in the TPC ($E/p$), as well as on the shape of the corresponding electromagnetic shower ($\sigma^{2}_{\rm{long}}$). The $\sigma^{2}_{\rm{long}}$ refers to the eigenvalues of the dispersion matrix of the shower shape ellipse calculated by using the energy distribution within the EMCal cluster~\cite{Awes:1992yp, ALICE:2017nce}.
As the shower from an electron is fully contained and accurately measured by the EMCal, the $E/p$ ratio is expected to be close to unity, while the $E/p$ ratio for hadrons is very small. Electron candidates were selected with $0.85 < E/p < 1.3$ and $0.1 < \sigma^{2}_{\rm{long}} < 0.7$.

Electrons from W$^\pm$- and Z$^0$-boson decays are well isolated from other particles, allowing for their identification through an isolation requirement. In this analysis, the isolation condition is implemented by summing the cluster energies in the EMCal around the electron candidate within a radius  $R = \sqrt{(\Delta\eta)^{2}+(\Delta\varphi)^{2}}< 0.3$. 
The isolation energy $E_{\rm{iso}}$ is defined as a relative quantity 
\begin{equation}
     E_{\rm{iso}} = \left(\sum_{R_i<0.3} E_{\text{i,clus}}-E_{\rm{e}}\right)/E_{\rm{e}},
\end{equation}
where $E_{\rm{e}}$ represents the cluster energy for the electron candidate. A criterion of $E_{\textrm{iso}}<$ 0.1 is applied to distinguish electrons originating from W$^\pm$ and Z$^0$ bosons. Additionally, at most one charged particle in the cone ($N_{\textrm{R}<0.3}$) is required in order to improve the identification.

\begin{figure}[htb]
       \centering
     \includegraphics[width = 0.7\textwidth]{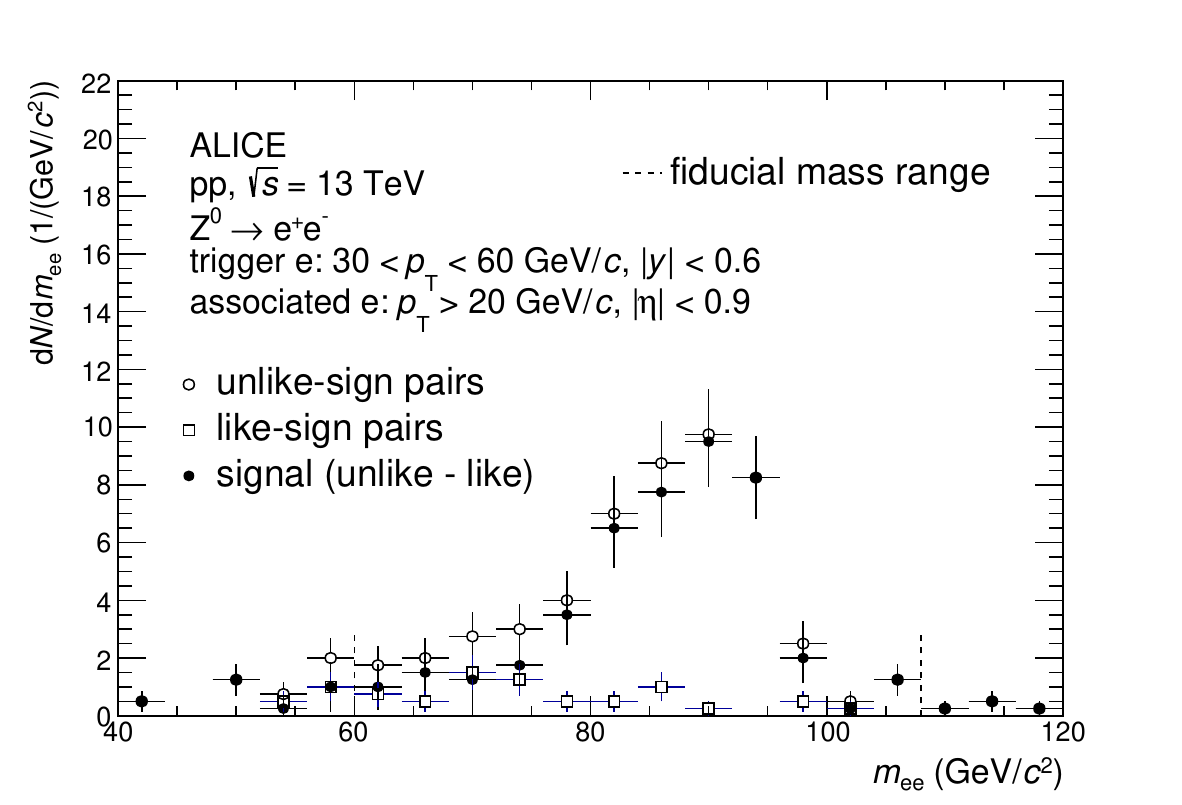}
     \caption{Invariant-mass distribution of ${\rm e^{+}e^{-}}$ after background subtraction in pp collisions at $\sqrt s = 13$~TeV. The like-sign (open squares) and unlike-sign (open circles) distributions are also shown. The dashed lines indicate the fiducial mass interval used in this analysis. The vertical lines on the data points are the statistical uncertainties.}
    \label{fig:Zmass}
\end{figure}

Z$^{0}$ bosons are reconstructed from the invariant mass of unlike-sign electrons in the range 60 $<m_{\rm ee}<$ 108 GeV/$c^{2}$. One electron is required to fulfill the electron identification and isolation criteria described above, as well as the fiducial condition 30 $<p_{\textrm{T}}<$ 60 GeV/$c$ and $|y|<$ 0.6, whereas the other one must satisfy loose identification criteria with $p_{\textrm{T}}>$ 20 GeV/$c$ in the TPC acceptance ($|\eta| <$ 0.9) and $-3 < n_\sigma^{\rm TPC} < 3$, without isolation cuts due to the limited acceptance of the EMCal. The background contamination is estimated based on the invariant mass of like-sign pairs.
The invariant-mass distribution after subtracting the background is shown in Fig.~\ref{fig:Zmass} together with the distributions for unlike-sign and like-sign pairs. The resulting number of Z$^0$-boson candidates is then corrected for the efficiencies related to tracking, electron identification, isolation criteria and for finding the associated electron, and geometrical acceptance.  
The total reconstruction efficiency within the fiducial acceptance is approximately 6\% as evaluated via Monte Carlo simulations using PYTHIA6~\cite{Sjostrand:2006za} for Z$^0$ particle generation and GEANT3~\cite{Brun:1082634} for detector propagation.

\begin{figure}[htb]
    \begin{minipage}{0.5\hsize}
    \begin{center}
    \includegraphics[width = 1.\textwidth]{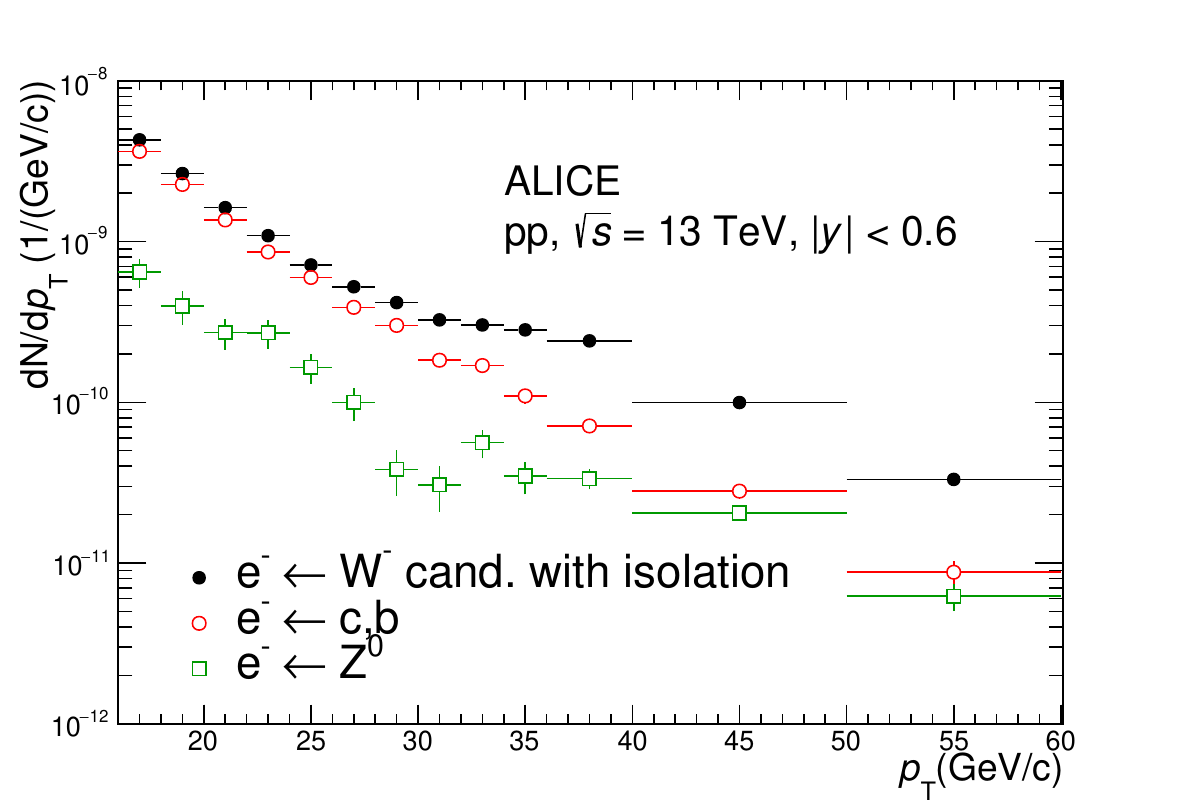}
    \end{center}
    \end{minipage}
    \begin{minipage}{0.5\hsize}
    \begin{center}
    \includegraphics[width = 1.\textwidth]{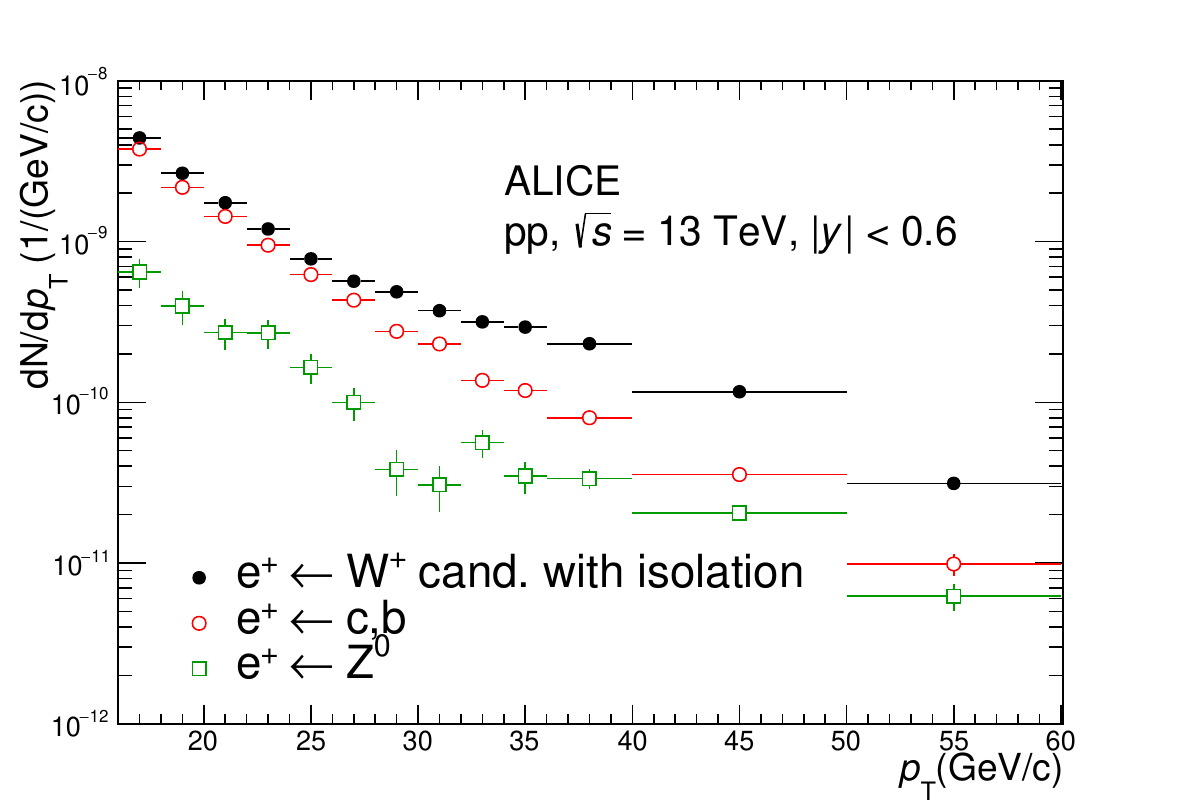}
    \end{center}
    \end{minipage}    
    \caption{Raw $p_{\textrm{T}}$ distribution for e$^{-} \leftarrow \textrm{W}^{-}$ (left) and e$^{+} \leftarrow \textrm{W}^{+}$ (right) candidates after the isolation conditions together with the estimated background from heavy-flavour hadron decays and Z$^0$ decays in pp collisions at $\sqrt{s}=$ 13 TeV. The symbols are placed at the centre of the $p_{\rm T}$ interval. The statistical uncertainties are represented by vertical bars.}
    \label{fig:Raw_ele}
\end{figure}

In the W$^\pm$-boson analysis, the remaining hadron contamination in the electron sample after the isolation conditions is estimated with a data-driven approach using the measured $E/p$ of hadrons. The background shape is obtained from the measurement of the $E/p$ distribution by selecting $n_{\sigma}^{\textrm{TPC}}<-2.5$ where the hadron contribution dominates. Then, the $E/p$ distribution of the background is normalized to match the distribution of electron candidates in 0.4 $< E/p <$ 0.7, where the hadron component is dominant. The purity is about 12 \%, and the hadron contribution is statistically subtracted from the sample.
After removing the hadron contamination, the residual background electrons fulfilling the isolation criteria originate from the decays of heavy-flavour hadrons (e $\leftarrow {\rm c, b}$) and Z$^0$. The raw yield of electrons from W$^\pm$ bosons after applying the criteria for isolation and the ones of background electrons are shown in Fig.~\ref{fig:Raw_ele}. The contribution of electrons from Z$^0$ is estimated experimentally by looking for the partner decay electron fulfilling the condition $60 < m_{\rm ee} < 108$~GeV/$c^2$. 
The obtained spectra are corrected for the probability of finding the partner electrons due to the limited acceptance and efficiency of the detectors. The contribution of electrons from Z$^{0}$ is statistically subtracted from the $p_{\textrm{T}}$ distribution of electrons from W$^{\pm}$-boson decays.
The impact of heavy-flavour hadron decay electrons is assessed using a data-driven approach. These electrons, emerging from charm- and beauty-quark fragments, generally show less isolation and exhibit higher $E_{\textrm{iso}}$ values. In this approach, they are selected by requiring $E_{\textrm{iso}} > 0.2$. 
The corresponding $p_{\textrm{T}}$ distribution is scaled to reproduce the $p_{\textrm{T}}$ distribution for electron candidates from W$^\pm$ bosons after removing the electrons from Z$^0$ in the range 16 $<p_{\textrm{T}}<$ 20 GeV/$c$, where the contribution of electrons from heavy-flavour hadron decays is the dominant source in the electron sample. The scaled spectrum is statistically subtracted from the $p_{\textrm{T}}$ distribution of electrons from W$^{\pm}$-boson decays.
The feed-down from top quarks is estimated using a POWHEG simulation~\cite{Alioli:2008gx}. The contribution is found to be smaller than 1\% and is neglected in the following. A contribution from electrons resulting from photon conversion in the detector material and Dalitz decays of neutral mesons (${\rm e}_{\rm \gamma}$) is also present. 
Such electrons are identified by requiring the mass of electron pairs $m_{{\rm ee}}$ to be $\sim 0$~\cite{ALICE:2015zhm}. A maximum contribution of 3\% is found for 
$p_{\textrm{T}}< 40$~GeV/$c$, and the contribution is negligible at higher $p_{\textrm{T}}$.
This contribution is not subtracted but treated as a systematic uncertainty. The number of identified electrons from W$^\pm$-boson decays is corrected for the reconstruction efficiency to calculate the production cross section. The reconstruction efficiency is evaluated with simulations of W$^\pm$ bosons generated with PYTHIA 6~\cite{Sjostrand:2006za} and propagated through the detector with GEANT3~\cite{Brun:1082634}. The total reconstruction efficiency, including tracking, electron identification, and isolation selections, is about 11\% in the analyzed kinematic range, and is found to be independent of the multiplicity class.

The yield of electrons from W$^\pm$-boson decays in the different multiplicity classes is obtained using the same selection criteria and background subtraction approach as for the multiplicity-integrated sample. Particles produced in association with at least one W$^{\pm}$ boson are selected from their azimuthal correlation with the 
$\rm W^\pm$ electron candidate ($\Delta \varphi_{\textrm{e}^{{\rm W-h}}}$). The associated particles are selected within a pseudorapidity range $|\eta|<$ 0.9, using the same track selection criteria as for the e$^{\pm} \leftarrow \textrm{W}^{\pm}$ reconstruction.
Furthermore, the particles are required to have a transverse momentum $p_{\textrm{T}}>$ 10 GeV/$c$ to suppress contributions from soft hadrons originating from the underlying event.
Simulation studies within the experimental acceptance show a strong azimuthal correlation between the W$^{\pm}$ boson and its decay electron, supporting the use of the electron direction as an indication for the boson direction in the associated hadron analysis.
The background associated hadrons due to misidentification of e$^{\pm} \leftarrow \textrm {W}^{\pm}$ are estimated and statistically subtracted from the $\Delta \varphi_{\textrm{e}^{{\rm W-h}}}$ distribution. The correlations between hadrons and e$\leftarrow$~c, b are experimentally measured, whereas the ones between hadrons and e $\leftarrow$ Z$^{0}$ are obtained from PYTHIA simulations. These correlation distributions are normalised by their expected fraction in the e$^{\pm} \leftarrow \textrm {W}^{\pm}$ sample.
Since the associated particles predominantly arise from recoil quarks and are distributed on the away-side in azimuth with respect to the e$^{\pm} \leftarrow \textrm {W}^{\pm}$ decays, the number of associated particles is determined by integrating over $\vert \Delta \varphi_{\rm e}^{{\rm W-h}} - \pi \vert < 2$.

\section{Systematic uncertainties}

\subsection{Systematic uncertainties on ${\rm Z^0}\rightarrow \textrm{e}^{+}\textrm{e}^{-}$}
The systematic uncertainty sources on the production cross section of $Z^0\rightarrow \textrm{e}^{+}\textrm{e}^{-}$ are summarised in Table~\ref{tb:Zsys}.
The uncertainties from the electron identification in the TPC ($n_{\sigma}^{\textrm{TPC}}$) and the EMCal ($E/p$ and $\sigma^{2}_{\rm{long}}$) are assessed by varying the selection criteria and quantifying the observed change on the production cross section. 
These variations also account for possible effects related to the track momentum resolution and tracking efficiency, as well as the EMCal energy resolution, since such detector effects directly influence the reconstructed electron kinematics and the electron identification.
The resulting systematic uncertainty is 6\% from the $n_{\sigma}^{\textrm{TPC}}$ selection, 3\% from the $E/p$ selection, and 2\% from the shower-shape selection. 
The uncertainty arising from the isolation selection, based on $E_{\textrm{iso}}$ and $N_{\textrm{R}<0.3}$, is evaluated through a variation of the selection criteria and found to be smaller than 1\%. 
The background contribution in the signal mass range (60 $<m_{{\rm ee}}<$ 108 GeV/$c^{2}$) depends on the selections applied to the electron pair. This is studied by reconstructing the invariant mass with and without a minimum track $p_{\textrm{T}}$ and a condition on the opening angle for electron pairs. A systematic uncertainty of about 2\% is associated with the background subtraction.
The effect of the selections on the $\eta$ and $n_{\sigma}^{\textrm{TPC}}$ ranges for the electrons fulfilling looser criteria is studied as well, resulting in additional uncertainties of 3\% and 1\%, respectively.
The reconstruction efficiency is possibly influenced by the $Z^0$ mass shape in the Monte Carlo sample. While the nominal efficiency is estimated using the PYTHIA 6 event generator, a systematic impact of the underlying invariant-mass shape is investigated by considering the invariant-mass distribution obtained by POWHEG. A difference of 5.5\% is observed and assigned as a systematic uncertainty on the reconstruction efficiency.
Additionally, there is a 3\% uncertainty on single tracks arising from the matching of reconstructed trajectories between the ITS and the TPC~\cite{ALICE:2023xiu}. The uncertainty being correlated between two electron tracks, a 6\% uncertainty is assigned on the cross section measurement. The various uncertainties are assumed to be uncorrelated, and the total uncertainty is determined by the quadratic sum of the sources, resulting in a 11.5\% uncertainty for the Z$^{0}$.
Finally, an additional systematic uncertainty of 5.2\% arising from the integrated luminosity is mentioned separately in the rest of the publication. The latter is obtained as the quadratic sum of the systematic uncertainties of the V0 trigger cross section (1.6\%)~\cite{ALICE-PUBLIC-2021-005}  and the normalisation of the EMCal-triggered events to the equivalent number of V0-triggered events (5\%).

\begin{table}[htb]
\caption{Summary of relative systematic uncertainties on the production cross section for Z$^0\rightarrow {\rm e^{+}e^{-}}$ in pp collisions at $\sqrt{s}=$ 13 TeV.}
\begin{center}
  \begin{tabular}{l c} \hline
    Source & Uncertainty\\ \hline 
    $n_{\sigma}^{\textrm{TPC}}$ & 6\%    \\
    EMCal $E/p$ & 3\%  \\
    EMCal $\sigma^{2}_{\rm{long}}$ & 2\%  \\
    isolation selection $E_{\textrm{iso}}$ &   $<$ 1\% \\ 
    isolation selection $N_{\textrm{R}<0.3}$ &   $<$ 1\% \\ 
    background subtraction & 2\%  \\ 
    $n_{\sigma}^{\textrm{TPC}}$ on pair electrons& 1\%    \\
    $\eta$ on pair electrons & 3\%    \\
    reconstruction efficiency (input mass shape) & 5.5\%  \\ 
    TPC--ITS matching & 6\%  \\ 
    integrated luminosity & 5.2 \% \\ \hline
    Total (without luminosity) &  11.5\%  \\ \hline
  \end{tabular}
      \end{center}
      \label{tb:Zsys}
\end{table}

\subsection{Systematic uncertainties on e$^{\pm} \leftarrow \textrm{W}^{\pm}$}

The systematic uncertainty sources affecting the measurement of the production cross section for e$^{\pm} \leftarrow \textrm{W}^{\pm}$ are summarised in Table~\ref{tb:Wsys}.
The systematic uncertainty is estimated separately for e$^{+} \leftarrow \textrm{W}^{+}$ and e$^{-} \leftarrow \textrm{W}^{-}$ decays.
The systematic uncertainty from the electron identification in the TPC ($n_{\sigma}^{\rm{TPC}}$) and the EMCal ($E/p$ and $\sigma^{2}_{\rm{long}}$) is evaluated by varying the selection ranges and is found to be 5\% for each of the uncertainties, for both e$^{+} \leftarrow \textrm{W}^{+}$ and e$^{-} \leftarrow \textrm{W}^{-}$ decays. 
The uncertainty from the isolation selection based on the energy ($E_{\textrm{iso}}$) is found to be negligible while the one on the number of tracks ($N_{\textrm{R} < 0.3}$) is 10\%.
In the $\textrm{W}^{\pm}$ analysis, the impact of the track and EMCal performance is effectively propagated to the measured cross sections through variations of the electron identification and isolation selections, in particular the $E/p$ and $E_{\textrm{iso}}$ requirements that depend on the tracking and the EMCal response.
The systematic uncertainty resulting from the subtraction of  electrons from heavy-flavour hadron decays includes contributions from the isolation-energy condition  to make the template and a scaling factor for matching the template to the data. The first contribution is estimated by varying the $E_{\textrm{iso}}$ selection range and can be neglected. The scaling factor for the template has a 4\% systematic uncertainty which is determined by varying the $p_{\textrm{T}}$ range where the templates are fitted to data. The resulting uncertainty on the production cross section varies from 1\% to 3\% in the range 30 $<p_{\textrm{T}}<$ 60 GeV/$c$.
The uncertainty from the subtraction of electrons from Z$^0$ decays is obtained by propagating the 11.5\% total uncertainty of the Z$^0$$\rightarrow \textrm{e}^{+}\textrm{e}^{-}$ production cross section measurement to the e$^{\pm} \leftarrow \textrm{W}^{\pm}$ cross section. 
The impact is found to be smaller than 1\%. The 3\% contribution of the e$_{\gamma}$ in the region $p_{\textrm{T}}<$ 40 GeV/$c$ is assigned as an additional uncertainty in this $p_{\textrm{T}}$ interval. Finally, the 3\% uncertainty in the track matching between the ITS and the TPC needs to be considered.
The various systematic uncertainties are assumed to be uncorrelated and their quadratic sum results in a total systematic uncertainty of 13\%. As for the Z$^0$ analysis, a 5.2\% systematic uncertainty on the integrated luminosity is treated separately. The uncertainties are found to be identical for both the W$^+$ and W$^-$ production cross sections. 

The systematic uncertainties for the multiplicity dependent analysis of the e$^{\pm} \leftarrow \textrm{W}^{\pm}$ are estimated by considering the same sources as the ones summarised in Table 2. The uncertainties for electron identification and isolation selections are multiplicity dependent, varying from about 0.5\% to 12\%. The uncertainty on the associated hadron yields results from the isolation selections and the heavy-flavour electron subtraction. It varies between about 3\% and 6\%, depending on the multiplicity class. 

\begin{table}[htb]
\caption{Sources of systematic uncertainties affecting the production cross section for e$^{\pm} \leftarrow \textrm{W}^{\pm}$ in pp collisions at $\sqrt{s}=$ 13 TeV. All the values correspond to the multiplicity-integrated measurement and the ranges reflect the $p_{\rm{T}}$ dependence.}
\begin{center}
  \begin{tabular}{l c} \hline
    Source & Uncertainty \\ \hline
    $n_{\sigma}^{\textrm{TPC}}$ & 5\%  \\
    EMCal ($E/p$ and $\sigma^{2}_{\rm{long}}$) &5\%  \\
    isolation selection $E_{\textrm{iso}}$ & negligible \\ 
    isolation selection $N_{\textrm{R}<0.3}$ & 10\%  \\ 
    e $\leftarrow {\rm c}, {\rm b}$ background subtraction & 1\% $\rightarrow$ 3\% \\
    Z${^0}$ $\rightarrow$ ${\rm e^+e^-}$ subtraction& $<$ 1\% \\  
    e$_{\rm \gamma}$ contamination & 3\% ($p_{\textrm{T}}<$ 40 GeV/$c$) \\ 
    TPC--ITS matching & 3\% \\ 
    integrated luminosity & 5.2 \% \\ \hline
    Total (without luminosity) & 13\% \\ \hline
    
  \end{tabular}
  \label{tb:Wsys}
  \end{center}
\end{table}

\section{Results}

\subsection{Z$^0$-boson production}
\begin{figure}[t]
    \centering
    \includegraphics[width = 1.0\textwidth]{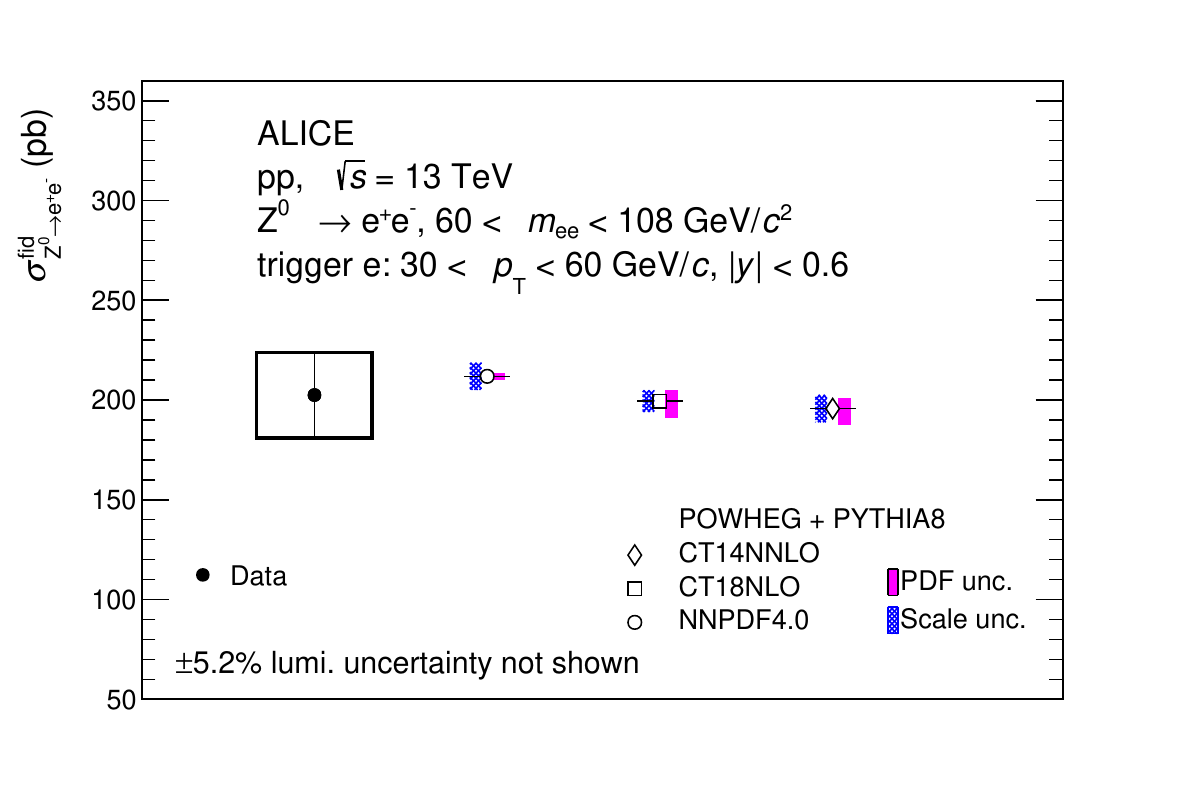}
    \caption{Production cross section of $Z^{0}$ bosons in pp collisions at $\sqrt{s}=$ 13 {\TeV} within the acceptance of the ALICE central barrel, compared with POWHEG calculations~\cite{Alioli:2008gx} performed with the CT14NNLO~\cite{Dulat:2015mca}, NNPDF40~\cite{NNPDF:2021njg} and CT18NLO~\cite{Yan:2022pzl} PDF sets. Statistical and systematic uncertainties of the measurement are shown as a vertical line and an open box, respectively.}
    \label{fig:CR_Zee}
\end{figure}

The production cross section for $\textrm{Z}^0 \rightarrow \textrm{e}^{+}\textrm{e}^{-}$ within 60 $<m_{{\rm ee}}<$ 108 GeV/$c^{2}$, with one electron fulfilling the requirements 30 $<p_{\textrm{T}}<$ 60 GeV/$c$ in $|y|<$ 0.6 in pp collisions at $\sqrt{s}=$ 13 TeV, is obtained as
\begin{equation}
\sigma^{\textrm{fid}}_{\textrm{Z}^0\rightarrow \textrm{e}^{+}\textrm{e}^{-}} = \frac{1}{\mathcal{L}}\frac{N_{\textrm{Z}^0\rightarrow \textrm{e}^{+}\textrm{e}^{-}}}{\epsilon},
\end{equation}
where $\mathcal{L}$ is the integrated luminosity, $N_{\textrm{Z}^0}\rightarrow \textrm{e}^{+}\textrm{e}^{-}$ is the number of reconstructed $Z^{0}$ bosons, and $\epsilon$ denotes the reconstruction efficiency.
The obtained result is shown in Fig.~\ref{fig:CR_Zee}.
The vertical  line represents the statistical uncertainty, and the box denotes the systematic uncertainty. 
The measured fiducial production cross section is
\begin{equation}
\sigma^{\textrm{fid}}_{{\textrm{Z}^0} \rightarrow \textrm{e}^{+}\textrm{e}^{-}} = 203 \pm 22\, (\textrm{stat.}) \pm 22\, (\textrm{sys.}) \pm 11\, (\textrm{lumi. uncert.})~\textrm{pb}.
\end{equation}
This value is obtained without correcting for the $\rm Z^0$ decay branching fraction into dielectrons.
A comparison with NLO pQCD calculations using different sets of parton distribution functions  is also displayed in Fig.~\ref{fig:CR_Zee}. 
The calculation is obtained using POWHEG to describe NLO vector-boson production~\cite{Alioli:2008gx} together with PYTHIA 8.2~\cite{Sjostrand:2014zea} for the parton shower and the CT14NNLO~\cite{Dulat:2015mca}, CT18NLO~\cite{Yan:2022pzl} and NNPDF4.0~\cite{NNPDF:2021njg} proton PDF sets.
CT14NNLO and CT18NLO provide PDFs at next-to-next-to-leading order (NNLO) and NLO level of accuracy in QCD calculations, respectively. NNPDF4.0 is based on a full global dataset and uses machine learning techniques. NNLO QCD calculations are employed and NLO electroweak corrections are considered. The calculations are performed using the same fiducial conditions as employed in the measurement. The mean value of the production cross section is represented as an open marker, and the scale uncertainties as well as the PDF uncertainty are depicted separately. 
The scale uncertainty is computed by varying the renormalisation factor ($\mu_{\rm R}$) and the factorisation factor ($\mu_{\rm F}$) relatively to the default choice $(\mu_{\rm R}, \mu_{\rm F})$ = (1, 1). The variations include $(\mu_{\rm R}, \mu_{\rm F})$ = (0.5, 0.5), (1, 0.5), (0.5, 1), (2, 2), (1, 2), and (2, 1).
The systematic uncertainties on the PDFs are computed following the prescriptions from Ref.~\cite{Harland-Lang:2014zoa}. 
The predicted production cross sections for the different PDF sets are consistent within PDF and scale uncertainties. The comparisons indicate that the measured Z$^0$-boson production cross section is well described by the theoretical calculations within experimental and theoretical uncertainties. 
The Z$^0$-boson production cross section in pp collisions at $\sqrt{s}=$ 13 TeV was also measured in the ATLAS, LHCb and CMS experiments. These results are found to be in agreement with theoretical calculations using the CT14, CT18 and NNPDF PDFs, among others~\cite{ATLAS:2016fij, LHCb:2016fbk, LHCb:2021huf,CMS:2024gzs}.

\subsection{W$^{\pm}$-boson production}
\subsubsection{Production cross section of e$^{\pm} \leftarrow \textrm{W}^{\pm}$}
\begin{figure}[htb]
    \begin{minipage}{0.33\hsize}
    \begin{center}
    \includegraphics[width = 1.1\textwidth]{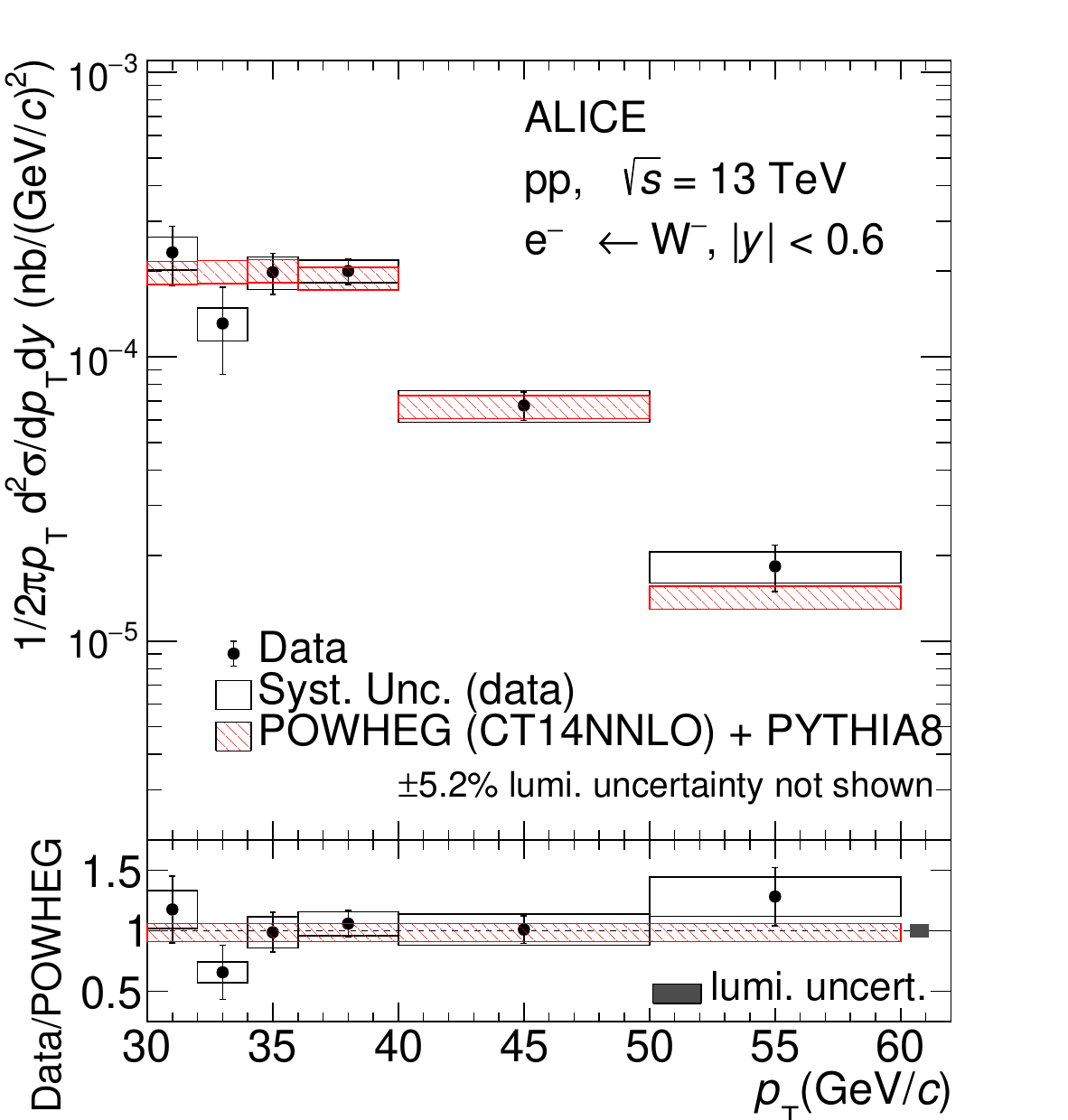}
    \end{center}
    \end{minipage}
    \begin{minipage}{0.33\hsize}
    \begin{center}
    \includegraphics[width = 1.1\textwidth]{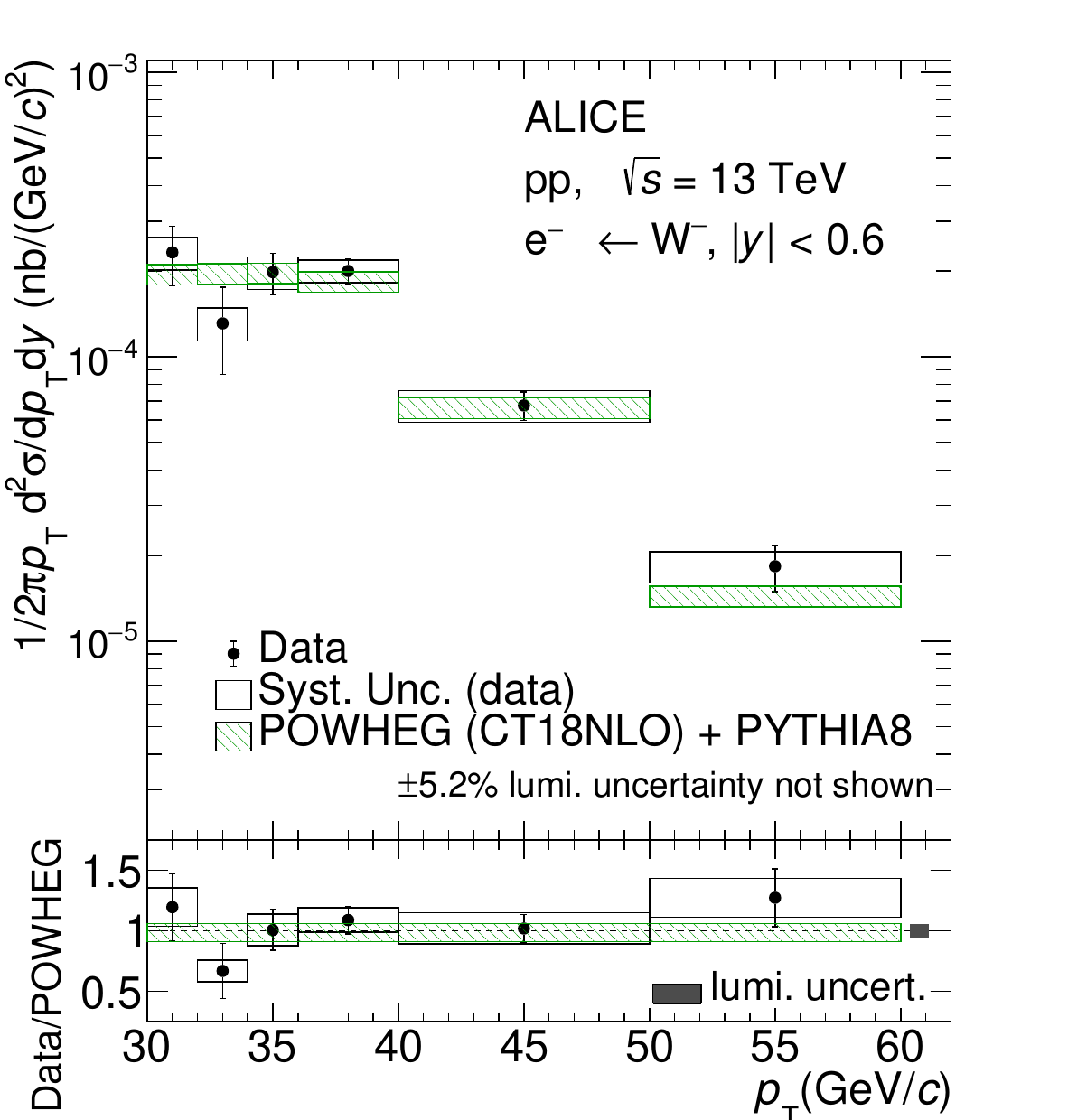}
    \end{center}
    \end{minipage}    
    \begin{minipage}{0.33\hsize}
    \begin{center}
    \includegraphics[width = 1.1\textwidth]{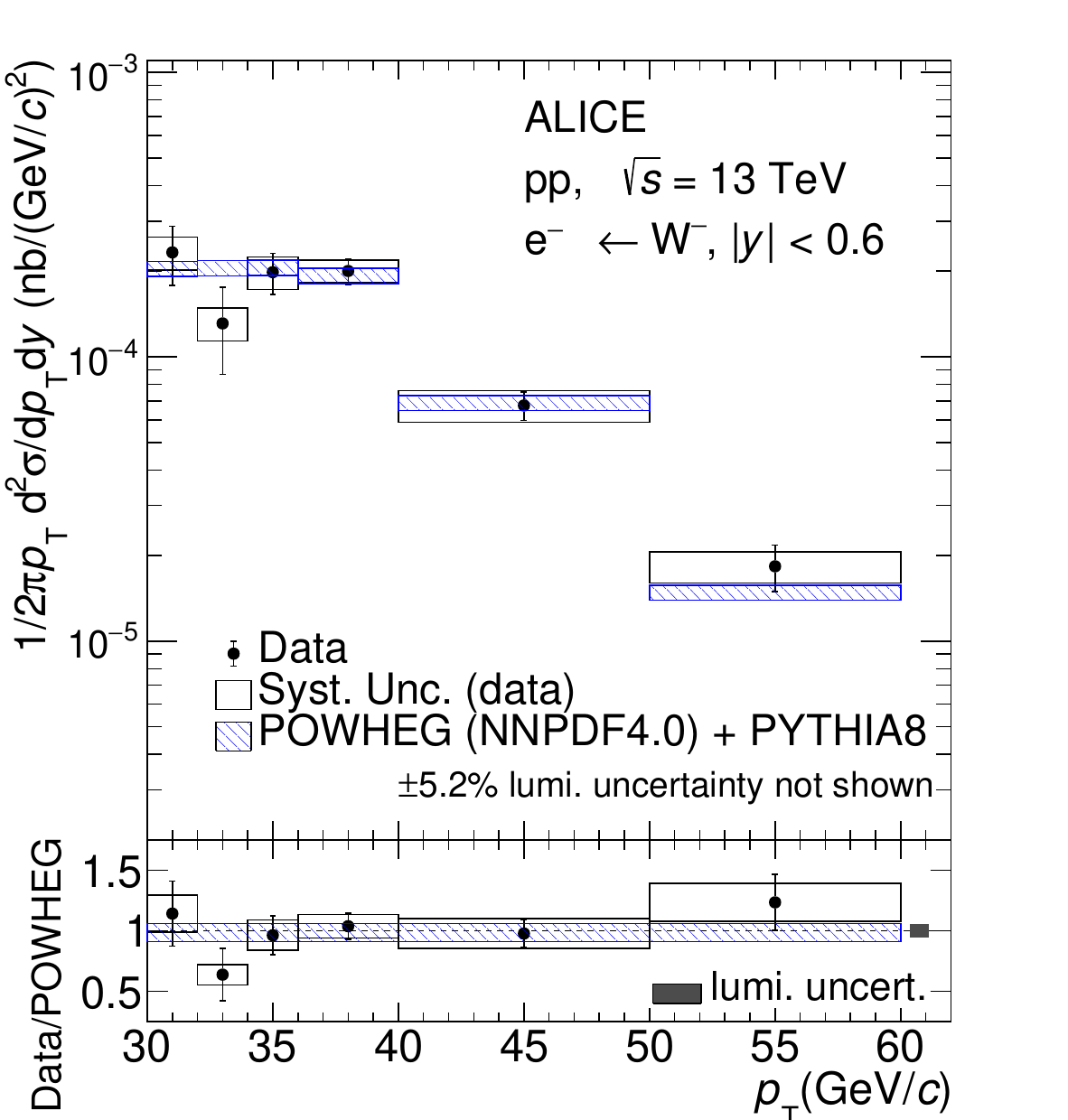}
    \end{center}
    \end{minipage}
    
    \caption{Production cross section of electrons from W$^{-}$ decays as a function of $p_{\rm T}$ at midrapidity in pp collisions at $\sqrt{s}=$13 {\TeV} compared with POWHEG calculations~\cite{Alioli:2008gx} using the CT14NNLO~\cite{Dulat:2015mca} (left), CT18NLO~\cite{Yan:2022pzl} (middle) and NNPDF4.0~\cite{NNPDF:2021njg} (right) PDF sets. The symbols are placed at the centre of the $p_{\rm T}$ interval. The statistical and systematic uncertainties of the measurement are shown as vertical lines and open boxes, respectively. The uncertainties of the theoretical calculations are shown as shaded bands. The ratio of the data to the POWHEG calculations is shown in the lower panels.}
    \label{fig:crptWe}
\end{figure}

\begin{figure}[htb]
    \begin{minipage}{0.33\hsize}
    \begin{center}
    \includegraphics[width = 1.1\textwidth]{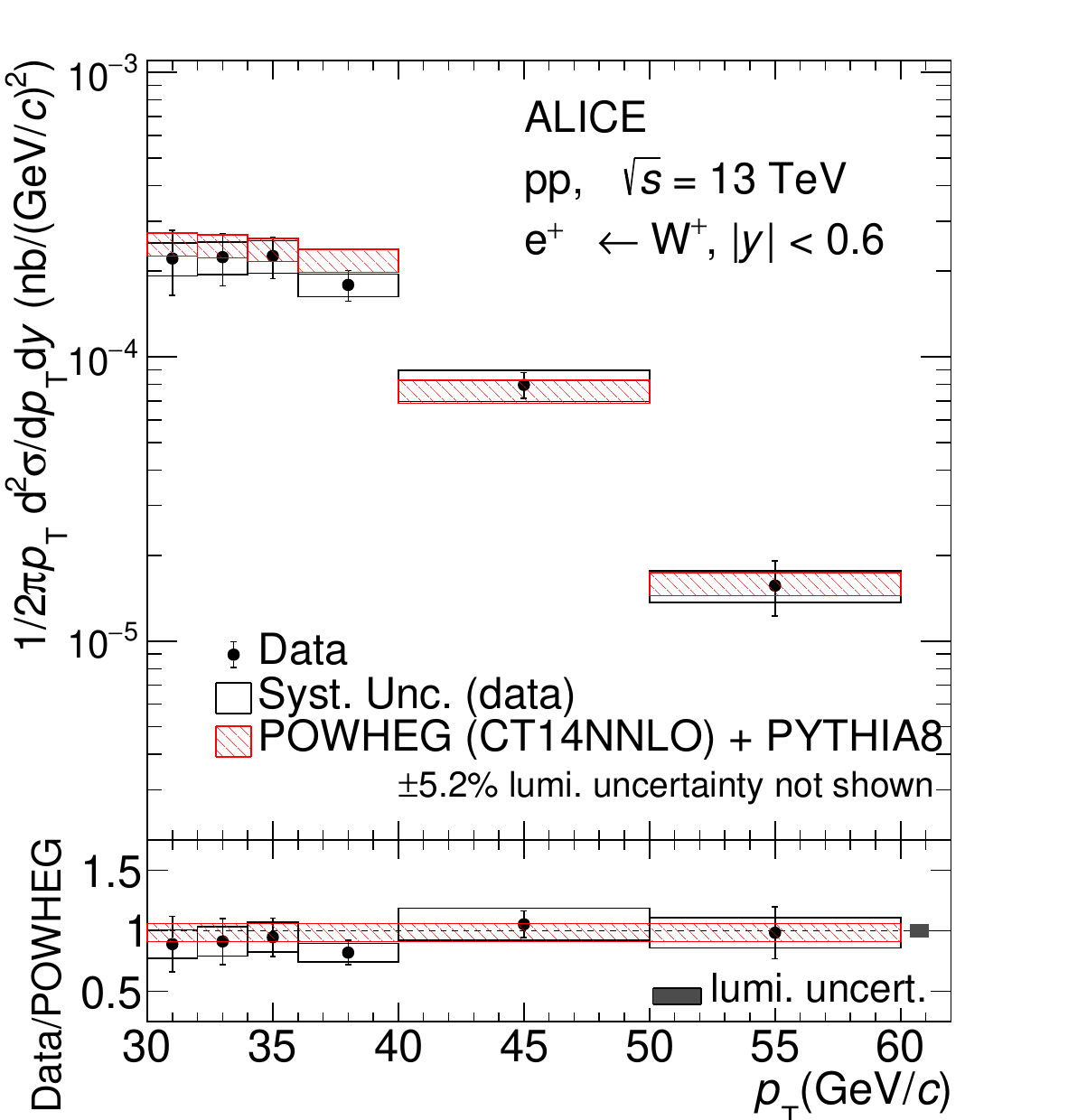}
    \end{center}
    \end{minipage}
    \begin{minipage}{0.33\hsize}
    \begin{center}
    \includegraphics[width = 1.1\textwidth]{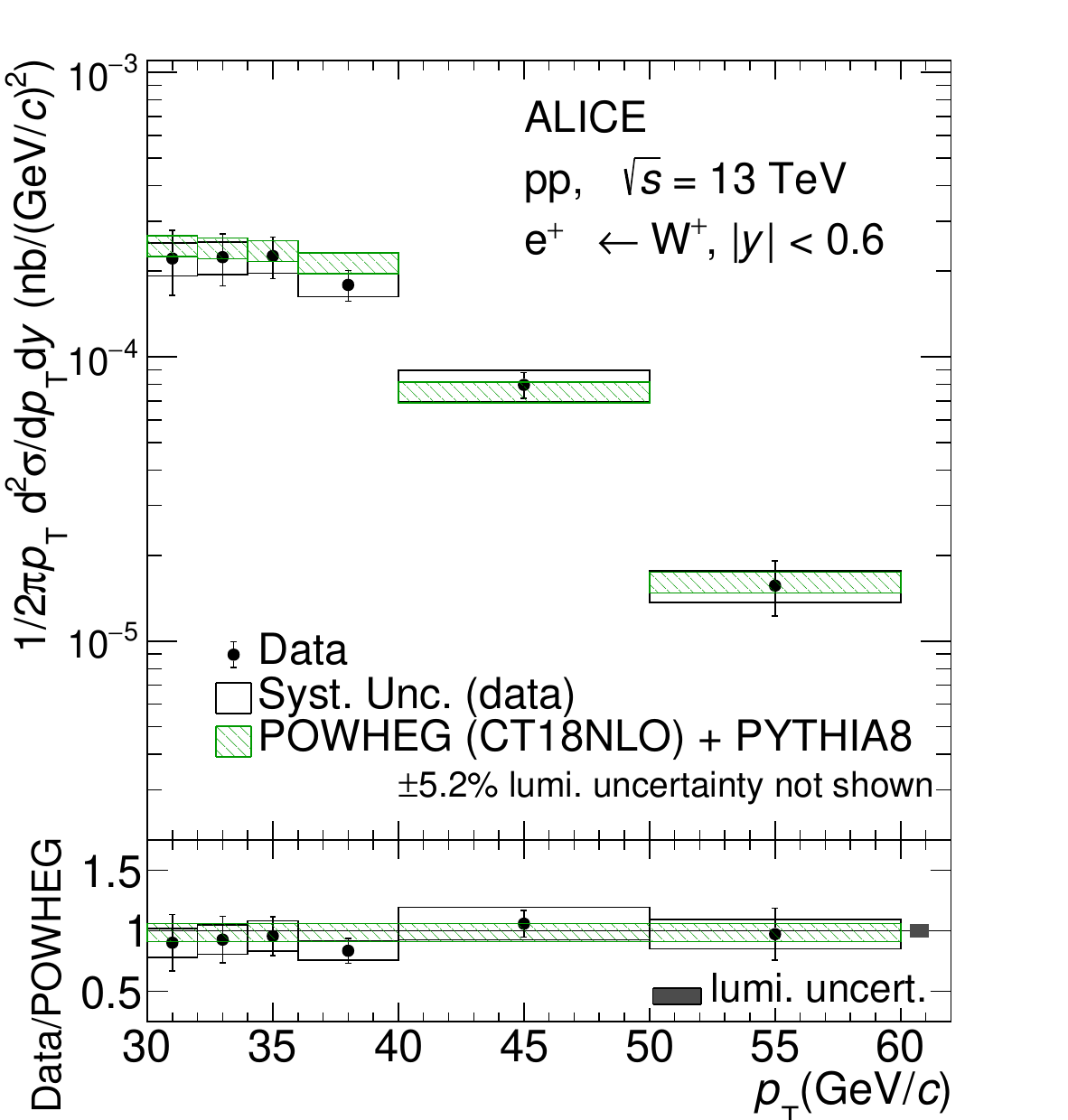}
    \end{center}
    \end{minipage}    
    \begin{minipage}{0.33\hsize}
    \begin{center}
    \includegraphics[width = 1.1\textwidth]{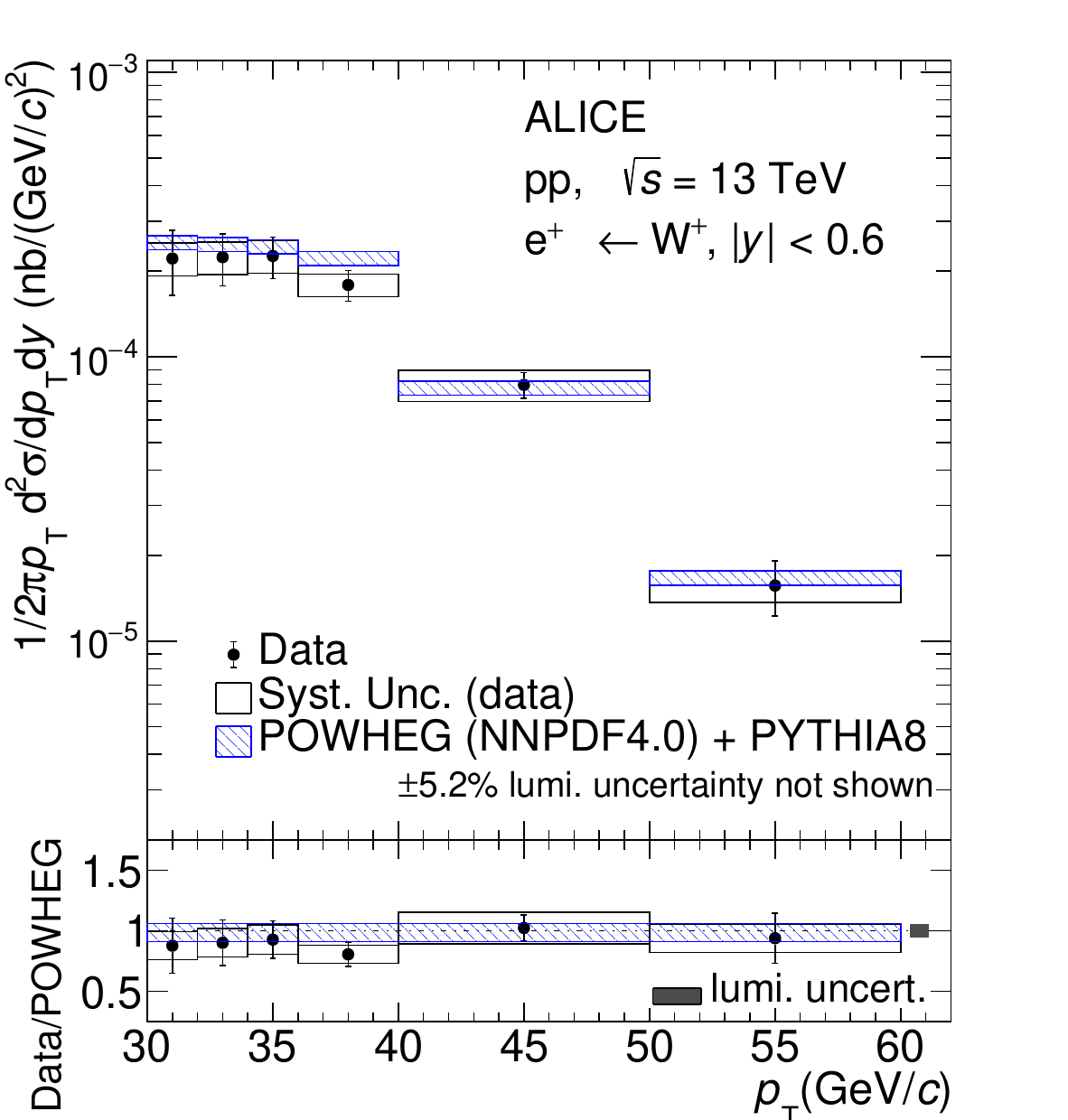}
    \end{center}
    \end{minipage}
    
    \caption{Production cross section of positrons from W$^{+}$decays as a function of $p_{\rm T}$ at midrapidity in pp collisions at $\sqrt{s}=$13 {\TeV} compared with POWHEG calculations~\cite{Alioli:2008gx} using the CT14NNLO~\cite{Dulat:2015mca} (left), CT18NLO~\cite{Yan:2022pzl} (middle) and NNPDF4.0~\cite{NNPDF:2021njg} (right) PDF sets. The symbols are placed at the centre of the $p_{\rm T}$ interval. The statistical and systematic uncertainties of the measurement are shown as vertical lines and open boxes, respectively. The uncertainties of the theoretical calculations are shown as shaded bands. The ratio of the data to the POWHEG calculations is shown in the lower panels.}
    \label{fig:crptWp}
\end{figure}

The $p_{\textrm{T}}$-differential production cross section for e$^{\pm} \leftarrow \textrm{W}^{\pm}$ in $\vert y \vert <$ 0.6, without correction for the W-to-electron branching ratio, is calculated as

\begin{equation}
\frac{1}{2\pi p_{\textrm{T}}}\frac{\textrm{d}^{2}\sigma_{\textrm{e}^{\pm}\leftarrow \textrm{W}^{\pm}}}{\textrm{d}p_{\textrm{T}}\textrm{d}y} = \frac{1}{\mathcal{L}}\frac{1}{2\pi p_{\textrm{T}}}\frac{N_{\textrm{e}^{\pm}\leftarrow \textrm{W}^{\pm}}}{\Delta p_{\textrm{T}}\Delta y}\frac{1}{\epsilon},
\end{equation}
where $\mathcal{L}$ is the integrated luminosity, $N_{\textrm{e}^{\pm}\leftarrow \textrm{W}^{\pm}}$ is the number of raw electrons (positrons) from W$^\textrm{-}$ (W$^\textrm{+}$) identified by applying the selections discussed in Section~3, and $\epsilon$ denotes the reconstruction efficiency.
The $p_{\textrm{T}}$-differential production cross section of e$^{-} \leftarrow \textrm{W}^{-}$ and e$^{+} \leftarrow \textrm{W}^{+}$ at midrapidity ($|y|<$ 0.6) in pp collisions at $\sqrt{s}=$ 13 TeV is shown in Figs.~\ref{fig:crptWe} and ~\ref{fig:crptWp}, respectively. The statistical uncertainties are indicated by the vertical lines, and the systematic uncertainties are represented as open boxes. The results are compared with theoretical calculations based on the POWHEG model used together with PYTHIA 8 for the parton shower and the CT14NNLO~\cite{Dulat:2015mca}, CT18NLO~\cite{Yan:2022pzl} and NNPDF4.0~\cite{NNPDF:2021njg} proton PDF sets. The calculations for e$^{-} \leftarrow \textrm{W}^{-}$ and e$^{+} \leftarrow \textrm{W}^{+}$ are illustrated as coloured boxes, encompassing the uncertainty from the PDFs and scale uncertainties ($\mu_{\textrm{F}}$ and $\mu_{\textrm{R}}$).
The systematic uncertainties are estimated in a similar way as for the Z$^0 \rightarrow \textrm{e}^{+}\textrm{e}^{-}$ predictions. 
The ratios of the measurements with respect to the theoretical predictions are displayed in the bottom panels.
The measured production cross sections are well described by the theoretical calculations within the uncertainties, showing the ability of pQCD calculations at NLO to reproduce the  e$^{-} \leftarrow \textrm{W}^{-}$ and e$^{+} \leftarrow \textrm{W}^{+}$ production cross sections and their $p_{\textrm{T}}$ dependence. These results are in line with previous measurements of the W-boson production in pp collisions at $\sqrt{s}=$ 13 TeV, where the ATLAS and CMS experiments reported a good agreement between the data and theoretical calculations using various PDF sets, amongst which are the CT14, CT18 and NNPDF PDFs~\cite{ATLAS:2016fij,CMS:2024gzs}. 

\begin{figure}[htb]
    \centering
    \includegraphics[width = 0.7\textwidth]{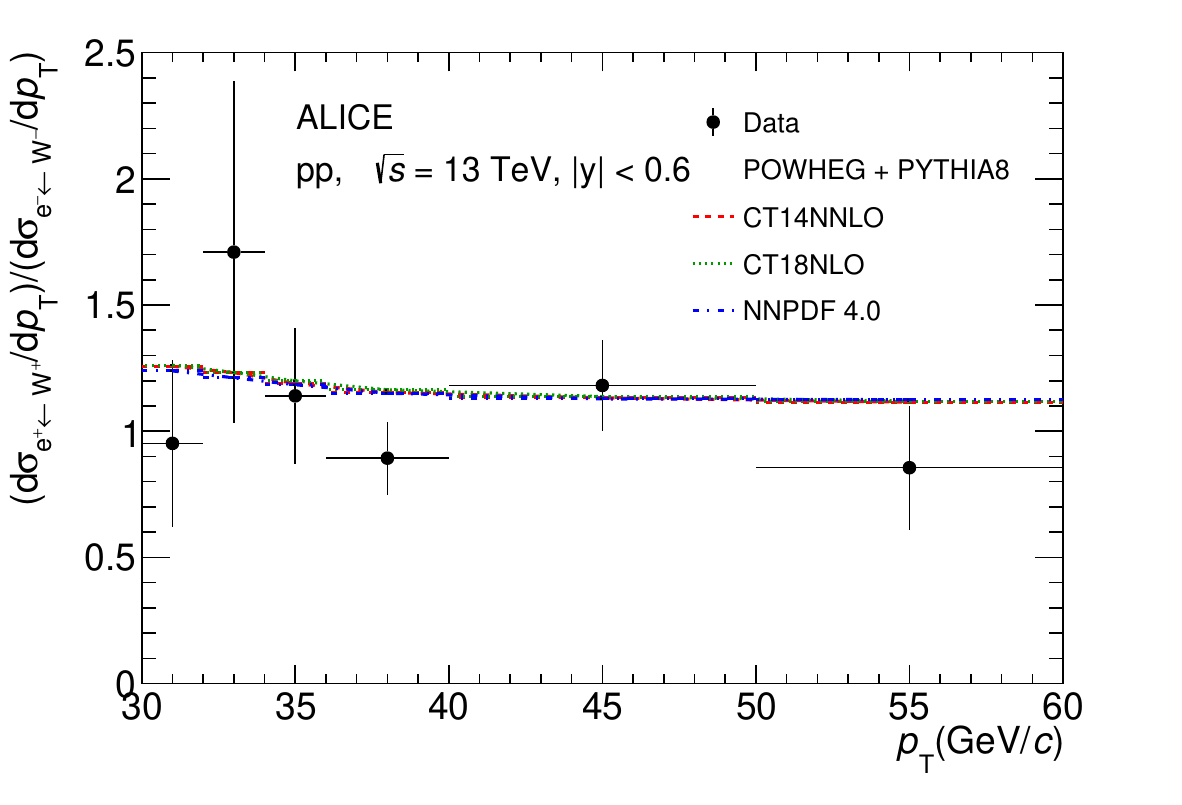}
    \caption{Ratio of the W$^+$ to W$^-$ production cross section as a function of $p_{\textrm{T}}$ at midrapidity in pp collisions at 
    $\sqrt{s}=$ 13 TeV. The symbols are placed at the centre of the $p_{\rm T}$ interval. The statistical uncertainties are represented by vertical bars.
    The curves with different colours are POWHEG calculations with the CT14NNLO~\cite{Dulat:2015mca}, CT18NLO~\cite{Yan:2022pzl} and NNPDF4.0~\cite{NNPDF:2021njg} PDF sets. Uncertainties on the calculations are assumed to cancel in the ratio.}
    \label{fig:eWratio}
\end{figure}

Utilising the measured production cross sections for e$^{-} \leftarrow \textrm{W}^{-}$ and e$^{+} \leftarrow \textrm{W}^{+}$, the production cross section ratio ($\displaystyle \frac{{\rm d}\sigma_{e^{+}\leftarrow \textrm{W}^{+}}}{{\rm d}p_{\rm T}}\bigg{/}\frac{{\rm d}\sigma_{e^{-}\leftarrow \textrm{W}^{-}}}{{\rm d}p_{\rm T}}$) is shown as a function of $p_{\textrm{T}}$ in Fig.~\ref{fig:eWratio}. Since the systematic uncertainties on e$^{-} \leftarrow \textrm{W}^{-}$ and e$^{+} \leftarrow \textrm{W}^{+}$ are similar and fully correlated, they are assumed to cancel in the ratio~\cite{ALICE:2022cxs}.  
The POWHEG predictions with the CT14NNLO, CT18NLO, and NNPDF4.0 PDF sets suggest that the ratio is above unity, indicating that the e$^{+} \leftarrow \textrm{W}^{+}$ production is larger than the e$^{-} \leftarrow \textrm{W}^{-}$ production. There is no significant difference in the production cross section ratio evaluated using  different PDFs. The measured ratio is dominated by statistical uncertainties,  making it challenging to determine the enhancement of e$^{+} \leftarrow \textrm{W}^{+}$ production compared to e$^{-} \leftarrow \textrm{W}^{-}$ production. 
The theoretical predictions are consistent with data within the uncertainties.

The $p_{\textrm{T}}$-integrated cross sections at $\vert y \vert < 0.6$ ($\textrm{d}\sigma / \textrm{dy}$) for e$^{-} \leftarrow \textrm{W}^{-}$ and e$^{+} \leftarrow \textrm{W}^{+}$ are determined by integrating over the electron (positron) $p_{\textrm{T}}$ in the interval $30 < p_{\rm T} < 60$ GeV/$c$. The measured production cross sections are 
\begin{equation}
\frac{\textrm{d}\sigma_{\textrm{e}^{-}\leftarrow \textrm{W}^{-}}}{\textrm{d}y} = 0.68 \pm 0.04\, (\textrm{stat.}) \pm 0.09\, (\textrm{sys.}) \pm 0.04\, (\textrm{lumi. uncert.})~\textrm{nb},
\end{equation}
\begin{equation}
\frac{\textrm{d}\sigma_{\textrm{e}^{+}\leftarrow \textrm{W}^{+}}}{\textrm{d}y}= 0.72 \pm 0.05\, (\textrm{stat.}) \pm 0.09\, (\textrm{sys.}) \pm 0.04\, (\textrm{lumi. uncert.})~\textrm{nb}.
\end{equation}

\begin{figure}[htb]
    \begin{minipage}{0.5\hsize}
    \begin{center}
    \includegraphics[width = 1.1\textwidth]{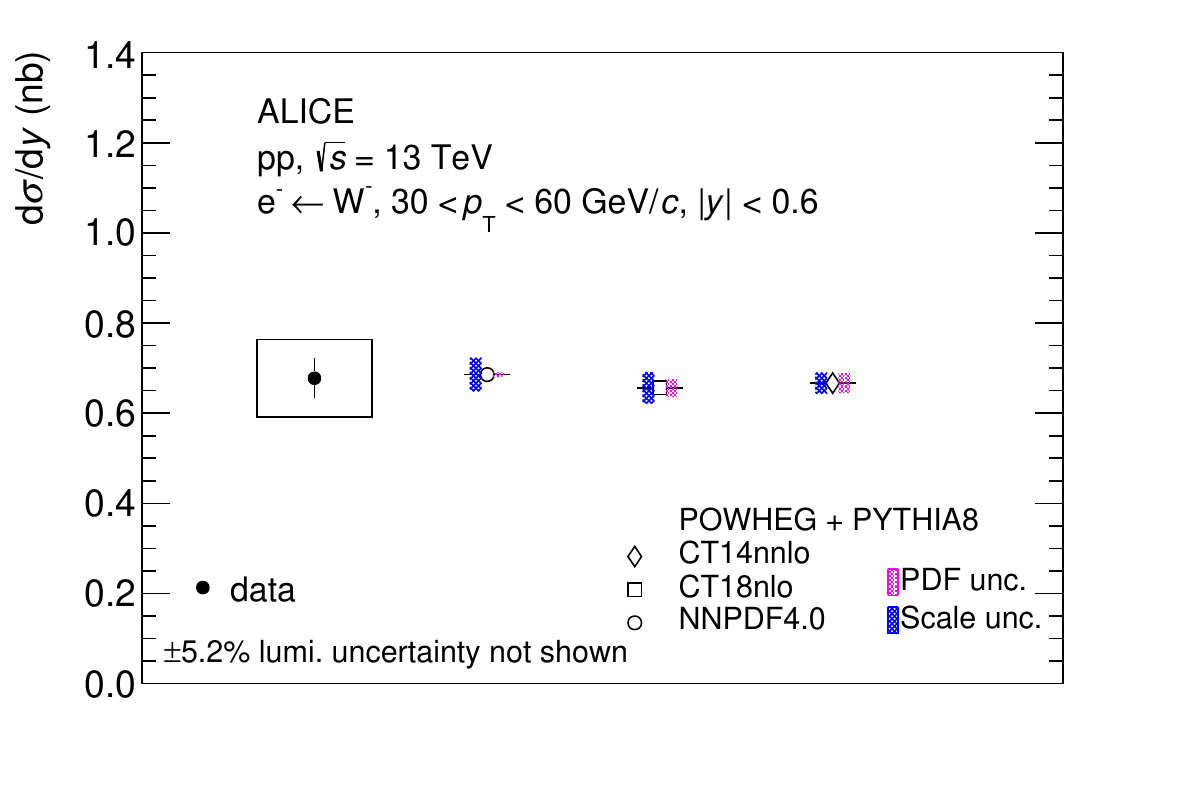}
    \end{center}
    \end{minipage}
    \begin{minipage}{0.5\hsize}
    \begin{center}
    \includegraphics[width = 1.1\textwidth]{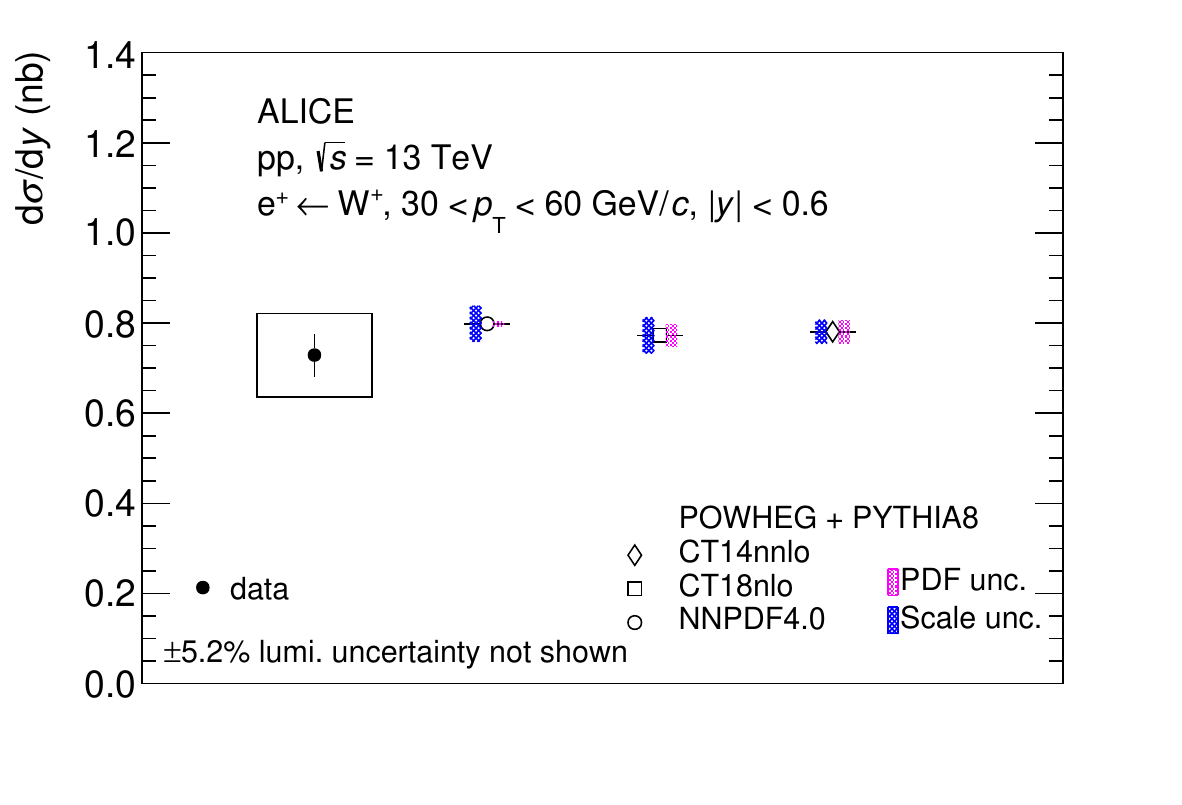}
    \end{center}
    \end{minipage}    
    \caption{Production cross section of electrons from W$^{-}$ (left) and positrons from W$^{+}$ (right) decays at midrapidity in pp collisions at $\sqrt{s}=$13 {\TeV}, compared with POWHEG calculations with the  CT14NNLO~\cite{Dulat:2015mca}, CT18NLO~\cite{Yan:2022pzl} and NNPDF4.0~\cite{NNPDF:2021njg} PDF sets. The production cross sections are obtained by integrating e$^{\pm} \leftarrow \textrm{W}^{\pm}$ in the interval $30 < p_{\rm T} < 60$ ~GeV/$c$. Statistical and systematic uncertainties of the measurement are shown as a vertical line and an open box, respectively.}
    \label{fig:IntegeW}
\end{figure}

The results are compared with POWHEG calculations using the CT14NNLO, CT18NLO and NNPDF4.0 PDF sets in Fig.~\ref{fig:IntegeW}. A good agreement is observed between the model calculations and the data within uncertainties.

\subsubsection{Multiplicity dependence of W${^\pm}$-boson and associated hadron production}

For the multiplicity-dependent measurement, the signal of electrons from W$^\pm$-boson decays ($N_{\textrm{e} \leftarrow \textrm{W}}$) is integrated in the electron $p_{\rm T}$ range 30 $<p_{\textrm{T}}<$ 60 GeV/$c$. Using the number of triggered events and the rejection factor in a given multiplicity class, the yield of electrons from W$^\pm$-boson decays is calculated as
\begin{equation}
\frac{\textrm{d}^{2}N^{\textrm{imult}}}{\textrm{d} p_{\textrm{T}}\textrm{d}y} = \frac{N_{\textrm{e} \leftarrow \textrm{W}}^{\textrm{imult}}}{N_{\textrm{evt, EG1}}^{\textrm{imult}}\times R^{\textrm{imult}} \times \Delta p_{\textrm{T}} \times \Delta y},
\end{equation}
where ``imult" indicates the multiplicity class, $N_{\textrm{evt, EG1}}^{\textrm{imult}}$ is the number of events in the EG1 trigger for a given multiplicity class, and $R^{\textrm{imult}}$ is the trigger rejection factor (see Section 2). The self-normalised yields are calculated as
\begin{equation}
\frac{\textrm{d}^{2}N}{\textrm{d}p_{T}\textrm{d}y}\bigg/\langle\frac{\textrm{d}^{2}N}{\textrm{d}p_{T}\textrm{d}y}\rangle = \frac{\textrm{d}^{2}N^{\textrm{imult}}_{e^\pm\leftarrow \textrm{W}^\pm}}{\textrm{d}p_{\textrm{T}}\textrm{d}y} \bigg/ \frac{\textrm{d}^{2}N^{\textrm{All}}_{{\rm e^\pm} \leftarrow \textrm{W}^\pm}}{\textrm{d}p_{\textrm{T}}\textrm{d}y},
\label{eq_mult}
\end{equation}
where $N^{\textrm{All}}_{{\rm e^\pm} \leftarrow \textrm{W}^\pm}$ represents the yield of e$^\pm$ $\leftarrow \textrm{W}^\pm$ in the integrated charged-particle multiplicity interval 0~$<N_{\textrm{tracklets}}<$ 200 in INEL $>$ 0 events. 
Here, $N_{\textrm{tracklets}}$ denotes the number of SPD tracklets measured within the acceptance $|\eta| <$ 1 on an event-by-event basis. The upper bound was chosen to safely encompass the full multiplicity distribution of the selected events.
The self-normalised yields of associated hadrons with $p_{\rm T} > 10$~GeV/$c$ for the various multiplicity classes are determined in a way similar 
to the one described in Eq.~(\ref{eq_mult}).

\begin{figure}[htb]
    \begin{minipage}{0.5\hsize}
    \begin{center}
    \includegraphics[width = 1.1\textwidth]{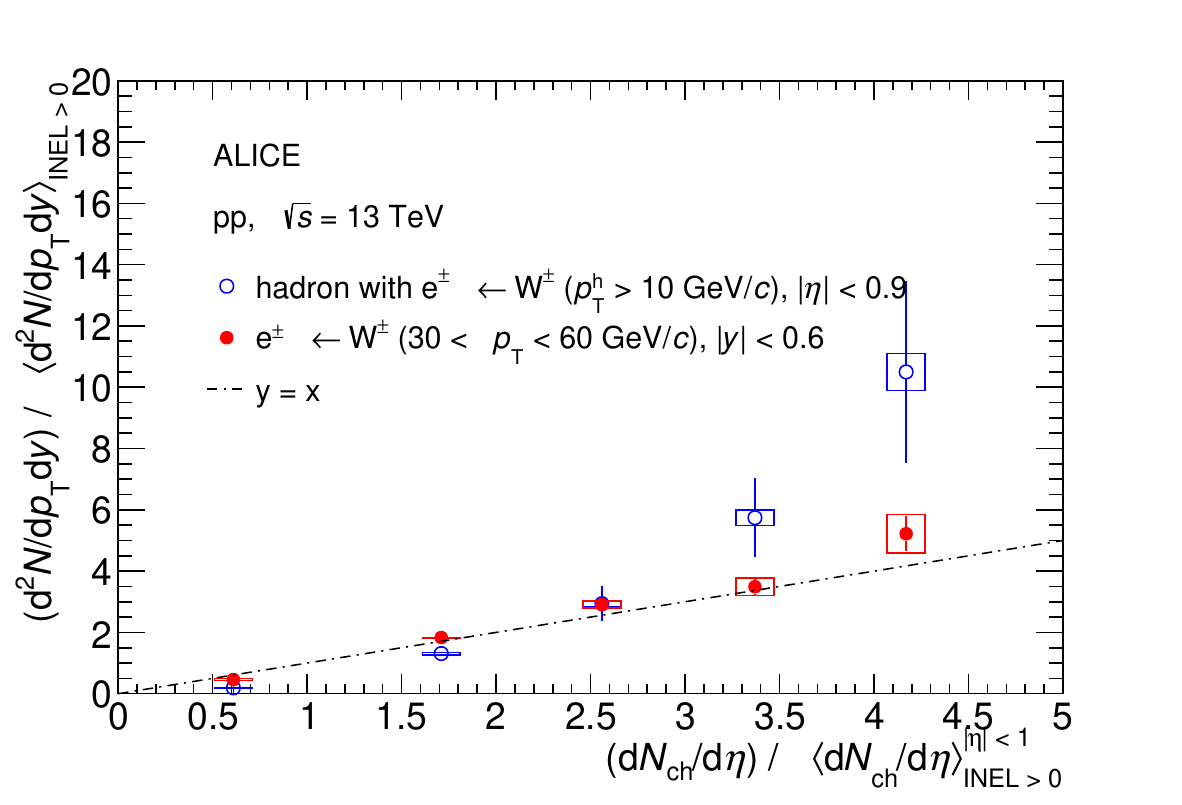}
    \end{center}
    \end{minipage}
    \begin{minipage}{0.5\hsize}
    \begin{center}
    \includegraphics[width = 1.1\textwidth]{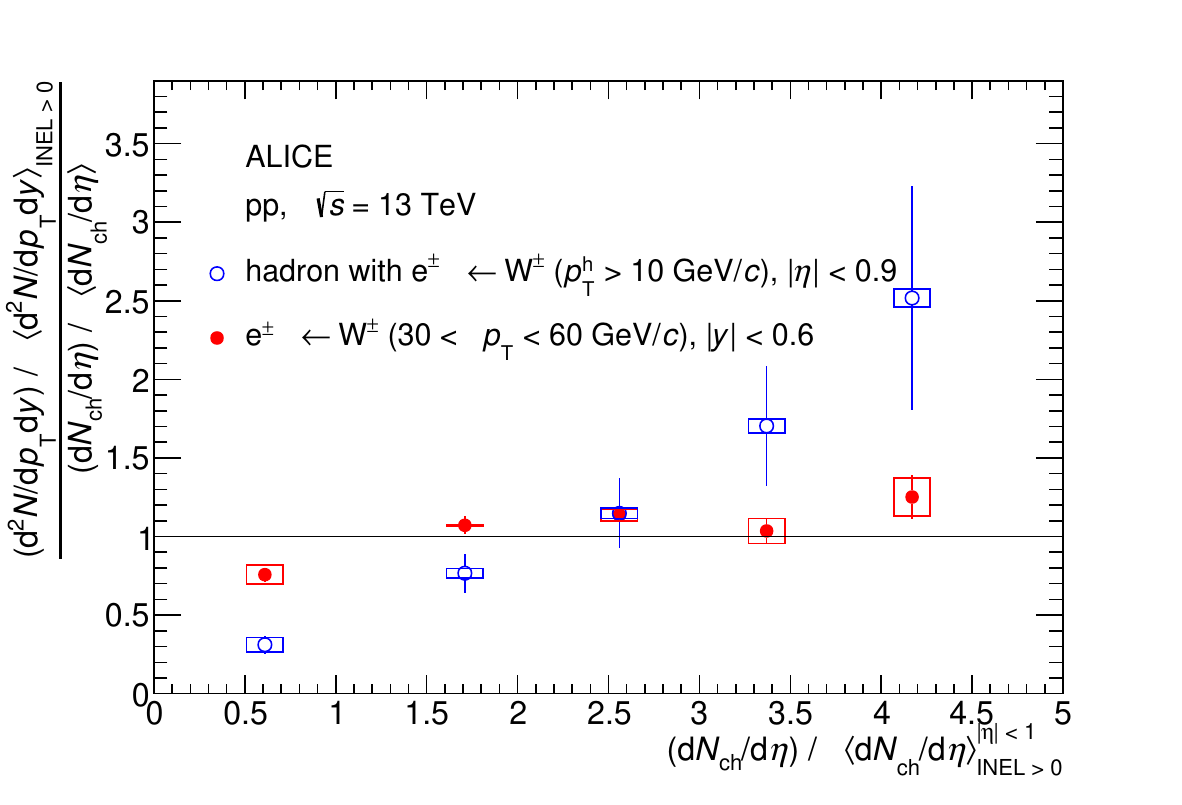}
    \end{center}
    \end{minipage}    
    \caption{Left: Self-normalised yields of e$^{\pm}$ $\leftarrow$ W$^{\pm}$ and associated hadrons as a function of the  normalised charged-particle pseudorapidity density. The linear behaviour $y=x$ is also shown as a dash-dotted line. The statistical and systematic uncertainties of the measurement are shown as vertical lines and open boxes, respectively. Right: Double ratio of the self-normalized yields of $e^{\pm} \leftarrow \textrm{W}^{\pm}$ to the self-normalised charged-particle multiplicity in pp collisions at $\sqrt{s}=$ 13 TeV.}
    \label{fig:ewmult}
\end{figure}

Figure~\ref{fig:ewmult} (left) shows the self-normalised yields of electrons from W$^{\pm}$-boson decays and charged hadrons in association with W$^\pm$ bosons as a function of the normalised charged-particle multiplicity at midrapidity in pp collisions at $\sqrt{s}=$ 13 {\TeV}. The dash--dotted line corresponds to a linear function with a slope of unity. The double ratio of the self-normalized yields of $e^{\pm} \leftarrow \textrm{W}^{\pm}$ to the self-normalised multiplicity in pp collisions at $\sqrt{s}=$ 13 TeV is also shown in the right panel.
The production of electrons from W$^{\pm}$ bosons increases with increasing charged-particle multiplicity and follows an approximately linear behaviour.
In contrast, the evolution of the associated hadron production as a function of the normalised charged-particle multiplicity tends to be faster-than-linear in particular for high charged-particle multiplicities. This enhancement in particle production beyond a linear dependence was observed by the ALICE collaboration for high--$p_{\rm{T}}$ charged particles~\cite{ALICE:2019dfi} and strange hadrons~\cite{ALICE:2019avo}, D mesons~\cite{ALICE:2015ikl}, electrons from heavy-flavour hadron decays~\cite{ALICE:2023xiu}, and J/$\psi (\rightarrow \textrm{e}^{+}\textrm{e}^{-})$~\cite{ALICE:2020msa} in pp collisions at midrapidity. 

\begin{figure}[htb]
    \begin{minipage}{0.5\hsize}
    \begin{center}
    \includegraphics[width = 1.1\textwidth]{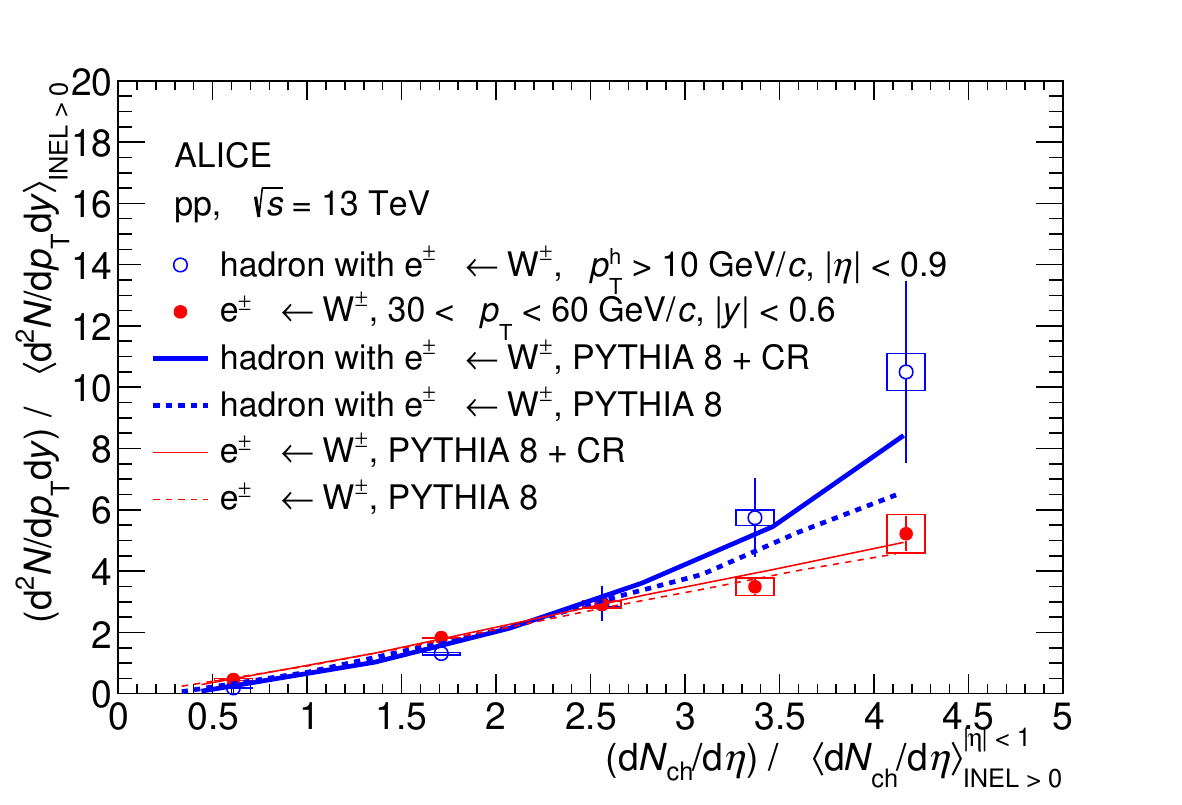}
    \end{center}
    \end{minipage}
    \begin{minipage}{0.5\hsize}
    \begin{center}
    \includegraphics[width = 1.1\textwidth]{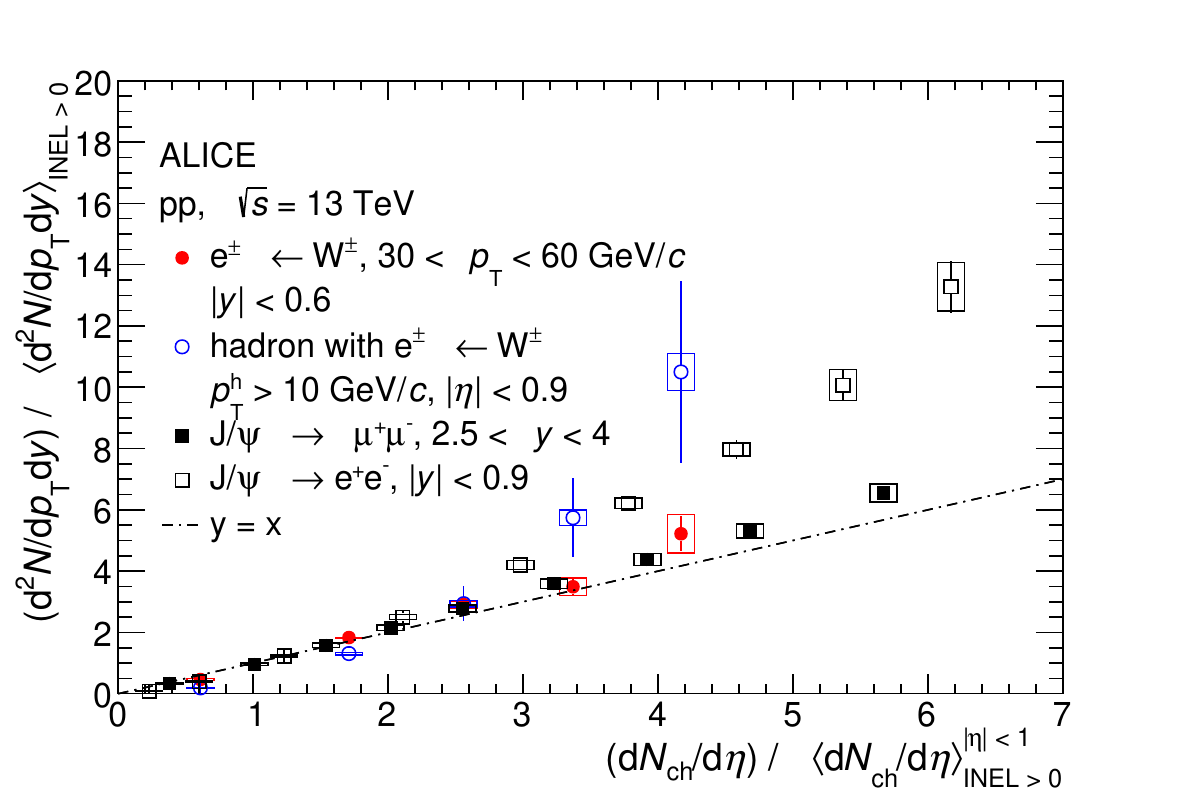}
    \end{center}
    \end{minipage}    
    \caption{Left: Self-normalised yields of e$^{\pm}$ $\leftarrow$ W$^{\pm}$ and associated hadrons as a function of the  normalised charged-particle pseudorapidity density at midrapidity in pp collisions at $\sqrt{s}=$ 13 TeV, compared with the expectations from PYTHIA 8~\cite{Sjostrand:2014zea} simulations including MPI and with/without CR effects. Right: self-normalised yields of e$^{\pm}$ $\leftarrow$ W$^{\pm}$ and associated hadrons compared with J/$\psi \rightarrow \textrm{e}^{+}\textrm{e}^{-}$ at midrapidity ($|y_{ee}|<$ 0.9)~\cite{ALICE:2020msa} and J/$\psi \rightarrow {\rm \mu^{+}\mu^{-}}$ at forward rapidity (2.5 $<y_{\mu\mu}<$ 4)~\cite{ALICE:2021zkd}, all yields being measured as a function of the self-normalised charged-particle pseudorapidity density at midrapidity. The statistical and systematic uncertainties of the measurements are shown as vertical lines and open boxes, respectively.}
    \label{fig:ewmult_comp}
\end{figure}

The observed trend has multiple possible interpretations.
In the colour-reconnection model, partons from different hard scatterings can connect through their colour charge to minimise the total string length. This process forms hadrons from partons involved in different hard scatterings, leading to an increase of the mean  transverse momentum with increasing charged-particle multiplicity~\cite{Sarma:2021qfr, OrtizVelasquez:2013ofg}.
W$^{\pm}$ bosons, being colourless, are not expected to be influenced by colour reconnections, resulting in an approximately linear relationship between production and multiplicity.
On the other hand, when partons associated with W$^{\pm}$ bosons contribute to the hadron formation through colour reconnection, there is an enhancement in hadron production compared to the linear prediction, especially in high-multiplicity events.
The results are compared in Fig.~\ref{fig:ewmult_comp} (left) with PYTHIA 8 (8.243) calculations~\cite{Sjostrand:2014zea} using the Monash 2013 tune and incorporating MPI and CR effects, which reproduce the charged-particle multiplicity distribution measured at the LHC~\cite{ATLAS:2017wln}.
The calculations indicate a linear trend for the multiplicity dependence of the production of e$^{\pm} \leftarrow \textrm{W}^{\pm}$ with and without colour reconnection. The result is consistent with the expectation that the production of W$^{\pm}$ bosons is not affected by colour reconnection. On the other hand, a faster-than-linear trend is obtained for the multiplicity dependence of the production of associated hadrons with and  without colour reconnection, although the trend is more pronounced when CR effects are included. Both calculations are consistent with the measurement within statistical and systematic uncertainties, although the data tend to favour the calculations with CR effects. Based on the studies with PYTHIA 8 discussed in Ref.~\cite{Weber:2018ddv}, another potential explanation of the observed trends is related to the autocorrelations between the measured hadrons and the charged-particle multiplicity. In Fig.~\ref{fig:ewmult_comp} (right), the multiplicity dependence of the production for electrons from W$^{\pm}$ bosons and charged hadrons in association with W$^\pm$ bosons is compared with J/$\psi$ $\rightarrow \mu^{+}\mu^{-}$ at forward rapidity (2.5 $< y_{\mu\mu}<$ 4) and J/$\psi$ $\rightarrow \textrm{e}^{+}\textrm{e}^{-}$ at midrapidity ($|y_{ee}|<$ 0.9)~\cite{ALICE:2020msa, ALICE:2021zkd}. The production of J/$\psi$ ($\rightarrow \textrm{e}^{+}\textrm{e}^{-}$) at midrapidity shows a faster-than-linear increase with the charged-particle multiplicity. 
In contrast, the production of J/$\psi$ $\rightarrow \mu^{+}\mu^{-}$  at forward rapidity does not exhibit a faster-than-linear trend as a function of the multiplicity. The production for J/$\psi$ ($\rightarrow \textrm{e}^{+}\textrm{e}^{-}$) at midrapidity is correlated with the charged-particle multiplicity due to the associated soft-particle production in events containing the J/$\psi$. Such autocorrelation effects could explain qualitatively the observed faster-than-linear trend. Furthermore, autocorrelation effects are expected to be less pronounced for the production of J/$\psi$ ($\rightarrow \mu^{+}\mu^{-}$) at forward rapidity, when the charged particle multiplicity
is measured at midrapidity. The correlation of W$^\pm$ bosons with the event multiplicity is expected to be small because of the isolated production. Therefore autocorrelation effects are likewise expected to be suppressed.
However, the hadrons produced in association with the W$^\pm$ bosons contribute to the determination of the event multiplicity, which can introduce additional autocorrelations.
The observed linear increase of the W$^{\pm}$ yield in contrast to the faster-than-linear increase of the associated hadron yield with increasing multiplicity tends to support the importance of autocorrelations when measuring the multiplicity dependence of charged particles. 

\section{Summary}
The production of electroweak bosons, W$^{\pm}$ and Z$^{0}$, is studied via the (di)electronic decay channels at midrapidity in pp collisions at $\sqrt{s}=$ 13 TeV. The production cross section of Z$^0$ in the fiducial kinematic range is measured by reconstructing the invariant mass of $\textrm{e}^{+}\textrm{e}^{-}$ pairs within 60 $< m_{{\rm ee}}<$ 108 GeV/$c^{2}$, where one electron is required to be in the ranges 30 $<p_{\textrm{T}}<$ 60 GeV/$c$ and $|y|<$ 0.6. The measured production cross section is compared with POWHEG calculations used together with the parton shower algorithm of the  PYTHIA 8 event generator with the CT14NNLO, CT18NLO, and NNPDF4.0 PDF sets. The calculations are found to be in good agreement with the experimental data.
The $p_{\textrm{T}}$-differential production cross sections of electrons from W$^{-}$ bosons and positrons from W$^{+}$ bosons are separately measured in the range 30 $<p_{\rm T} <$ 60 GeV/$c$ at midrapidity ($|y|<$ 0.6). The measured production cross sections are also compared with POWHEG using the CT14NNLO, CT18NLO, and NNPDF4.0 PDF sets, and a good agreement is observed. Furthermore, the production of W$^{\pm}$ is investigated as a function of the charged-particle multiplicity measured at midrapidity, and is found to 
exhibit an approximately linear trend. On the contrary, hadrons produced in association with W$^{\pm}$ present a faster-than-linear trend as a function of the charged-particle multiplicity. These results are well described by PYTHIA 8 calculations incorporating MPI effects and CR effects. These W and $\rm Z^0$ measurements will benefit from the large statistics collected with the upgraded ALICE detector during the  LHC Run 3 and Run 4.


\newenvironment{acknowledgement}{\relax}{\relax}
\begin{acknowledgement}
\section*{Acknowledgements}
\input{fa_2025-09-18_Opt_C.tex}
\end{acknowledgement}

\bibliographystyle{utphys}   
\bibliography{bibliography}

\newpage
\appendix

%
%

\section{The ALICE Collaboration}
\label{app:collab}
\input{Alice_Authorlist_2025-09-18_Opt_C.tex}  
\end{document}

%% file: commands.tex
%

\newcommand{\pp}           {pp\xspace}
\newcommand{\ppbar}        {\mbox{$\mathrm {p\overline{p}}$}\xspace}
\newcommand{\XeXe}         {\mbox{Xe--Xe}\xspace}
\newcommand{\PbPb}         {\mbox{Pb--Pb}\xspace}
\newcommand{\pA}           {\mbox{pA}\xspace}
\newcommand{\pPb}          {\mbox{p--Pb}\xspace}
\newcommand{\AuAu}         {\mbox{Au--Au}\xspace}
\newcommand{\dAu}          {\mbox{d--Au}\xspace}

\newcommand{\s}            {\ensuremath{\sqrt{s}}\xspace}
\newcommand{\snn}          {\ensuremath{\sqrt{s_{\mathrm{NN}}}}\xspace}
\newcommand{\pt}           {\ensuremath{p_{\rm T}}\xspace}
\newcommand{\meanpt}       {$\langle p_{\mathrm{T}}\rangle$\xspace}
\newcommand{\ycms}         {\ensuremath{y_{\rm CMS}}\xspace}
\newcommand{\ylab}         {\ensuremath{y_{\rm lab}}\xspace}
\newcommand{\etarange}[1]  {\mbox{$\left | \eta \right |~<~#1$}}
\newcommand{\yrange}[1]    {\mbox{$\left | y \right |~<~#1$}}
\newcommand{\dndy}         {\ensuremath{\mathrm{d}N_\mathrm{ch}/\mathrm{d}y}\xspace}
\newcommand{\dndeta}       {\ensuremath{\mathrm{d}N_\mathrm{ch}/\mathrm{d}\eta}\xspace}
\newcommand{\avdndeta}     {\ensuremath{\langle\dndeta\rangle}\xspace}
\newcommand{\dNdy}         {\ensuremath{\mathrm{d}N_\mathrm{ch}/\mathrm{d}y}\xspace}
\newcommand{\Npart}        {\ensuremath{N_\mathrm{part}}\xspace}
\newcommand{\Ncoll}        {\ensuremath{N_\mathrm{coll}}\xspace}
\newcommand{\dEdx}         {\ensuremath{\textrm{d}E/\textrm{d}x}\xspace}
\newcommand{\RpPb}         {\ensuremath{R_{\rm pPb}}\xspace}

\newcommand{\nineH}        {$\sqrt{s}~=~0.9$~Te\kern-.1emV\xspace}
\newcommand{\seven}        {$\sqrt{s}~=~7$~Te\kern-.1emV\xspace}
\newcommand{\twoH}         {$\sqrt{s}~=~0.2$~Te\kern-.1emV\xspace}
\newcommand{\twosevensix}  {$\sqrt{s}~=~2.76$~Te\kern-.1emV\xspace}
\newcommand{\five}         {$\sqrt{s}~=~5.02$~Te\kern-.1emV\xspace}
\newcommand{\twosevensixnn}{$\sqrt{s_{\mathrm{NN}}}~=~2.76$~Te\kern-.1emV\xspace}
\newcommand{\fivenn}       {$\sqrt{s_{\mathrm{NN}}}~=~5.02$~Te\kern-.1emV\xspace}
\newcommand{\LT}           {L{\'e}vy-Tsallis\xspace}
\newcommand{\GeVc}         {Ge\kern-.1emV/$c$\xspace}
\newcommand{\MeVc}         {Me\kern-.1emV/$c$\xspace}
\newcommand{\TeV}          {Te\kern-.1emV\xspace}
\newcommand{\GeV}          {Ge\kern-.1emV\xspace}
\newcommand{\MeV}          {Me\kern-.1emV\xspace}
\newcommand{\GeVmass}      {Ge\kern-.2emV/$c^2$\xspace}
\newcommand{\MeVmass}      {Me\kern-.2emV/$c^2$\xspace}
\newcommand{\lumi}         {\ensuremath{\mathcal{L}}\xspace}

\newcommand{\ITS}          {\rm{ITS}\xspace}
\newcommand{\TOF}          {\rm{TOF}\xspace}
\newcommand{\ZDC}          {\rm{ZDC}\xspace}
\newcommand{\ZDCs}         {\rm{ZDCs}\xspace}
\newcommand{\ZNA}          {\rm{ZNA}\xspace}
\newcommand{\ZNC}          {\rm{ZNC}\xspace}
\newcommand{\SPD}          {\rm{SPD}\xspace}
\newcommand{\SDD}          {\rm{SDD}\xspace}
\newcommand{\SSD}          {\rm{SSD}\xspace}
\newcommand{\TPC}          {\rm{TPC}\xspace}
\newcommand{\TRD}          {\rm{TRD}\xspace}
\newcommand{\VZERO}        {\rm{V0}\xspace}
\newcommand{\VZEROA}       {\rm{V0A}\xspace}
\newcommand{\VZEROC}       {\rm{V0C}\xspace}
\newcommand{\Vdecay} 	   {\ensuremath{V^{0}}\xspace}

\newcommand{\ee}           {\ensuremath{e^{+}e^{-}}} 
\newcommand{\pip}          {\ensuremath{\pi^{+}}\xspace}
\newcommand{\pim}          {\ensuremath{\pi^{-}}\xspace}
\newcommand{\kap}          {\ensuremath{\rm{K}^{+}}\xspace}
\newcommand{\kam}          {\ensuremath{\rm{K}^{-}}\xspace}
\newcommand{\pbar}         {\ensuremath{\rm\overline{p}}\xspace}
\newcommand{\kzero}        {\ensuremath{{\rm K}^{0}_{\rm{S}}}\xspace}
\newcommand{\lmb}          {\ensuremath{\Lambda}\xspace}
\newcommand{\almb}         {\ensuremath{\overline{\Lambda}}\xspace}
\newcommand{\Om}           {\ensuremath{\Omega^-}\xspace}
\newcommand{\Mo}           {\ensuremath{\overline{\Omega}^+}\xspace}
\newcommand{\X}            {\ensuremath{\Xi^-}\xspace}
\newcommand{\Ix}           {\ensuremath{\overline{\Xi}^+}\xspace}
\newcommand{\Xis}          {\ensuremath{\Xi^{\pm}}\xspace}
\newcommand{\Oms}          {\ensuremath{\Omega^{\pm}}\xspace}
\newcommand{\degree}       {\ensuremath{^{\rm o}}\xspace}

%% file: fa_2025-09-18_Opt_C.tex

The ALICE Collaboration would like to thank all its engineers and technicians for their invaluable contributions to the construction of the experiment and the CERN accelerator teams for the outstanding performance of the LHC complex.
The ALICE Collaboration gratefully acknowledges the resources and support provided by all Grid centres and the Worldwide LHC Computing Grid (WLCG) collaboration.
The ALICE Collaboration acknowledges the following funding agencies for their support in building and running the ALICE detector:
A. I. Alikhanyan National Science Laboratory (Yerevan Physics Institute) Foundation (ANSL), State Committee of Science and World Federation of Scientists (WFS), Armenia;
Austrian Academy of Sciences, Austrian Science Fund (FWF): [M 2467-N36] and Nationalstiftung f\"{u}r Forschung, Technologie und Entwicklung, Austria;
Ministry of Communications and High Technologies, National Nuclear Research Center, Azerbaijan;
Rede Nacional de Física de Altas Energias (Renafae), Financiadora de Estudos e Projetos (Finep), Funda\c{c}\~{a}o de Amparo \`{a} Pesquisa do Estado de S\~{a}o Paulo (FAPESP) and The Sao Paulo Research Foundation  (FAPESP), Brazil;
Bulgarian Ministry of Education and Science, within the National Roadmap for Research Infrastructures 2020-2027 (object CERN), Bulgaria;
Ministry of Education of China (MOEC) , Ministry of Science \& Technology of China (MSTC) and National Natural Science Foundation of China (NSFC), China;
Ministry of Science and Education and Croatian Science Foundation, Croatia;
Centro de Aplicaciones Tecnol\'{o}gicas y Desarrollo Nuclear (CEADEN), Cubaenerg\'{\i}a, Cuba;
Ministry of Education, Youth and Sports of the Czech Republic, Czech Republic;
The Danish Council for Independent Research | Natural Sciences, the VILLUM FONDEN and Danish National Research Foundation (DNRF), Denmark;
Helsinki Institute of Physics (HIP), Finland;
Commissariat \`{a} l'Energie Atomique (CEA) and Institut National de Physique Nucl\'{e}aire et de Physique des Particules (IN2P3) and Centre National de la Recherche Scientifique (CNRS), France;
Bundesministerium f\"{u}r Forschung, Technologie und Raumfahrt (BMFTR) and GSI Helmholtzzentrum f\"{u}r Schwerionenforschung GmbH, Germany;
General Secretariat for Research and Technology, Ministry of Education, Research and Religions, Greece;
National Research, Development and Innovation Office, Hungary;
Department of Atomic Energy Government of India (DAE), Department of Science and Technology, Government of India (DST), University Grants Commission, Government of India (UGC) and Council of Scientific and Industrial Research (CSIR), India;
National Research and Innovation Agency - BRIN, Indonesia;
Istituto Nazionale di Fisica Nucleare (INFN), Italy;
Japanese Ministry of Education, Culture, Sports, Science and Technology (MEXT) and Japan Society for the Promotion of Science (JSPS) KAKENHI, Japan;
Consejo Nacional de Ciencia (CONACYT) y Tecnolog\'{i}a, through Fondo de Cooperaci\'{o}n Internacional en Ciencia y Tecnolog\'{i}a (FONCICYT) and Direcci\'{o}n General de Asuntos del Personal Academico (DGAPA), Mexico;
Nederlandse Organisatie voor Wetenschappelijk Onderzoek (NWO), Netherlands;
The Research Council of Norway, Norway;
Pontificia Universidad Cat\'{o}lica del Per\'{u}, Peru;
Ministry of Science and Higher Education, National Science Centre and WUT ID-UB, Poland;
Korea Institute of Science and Technology Information and National Research Foundation of Korea (NRF), Republic of Korea;
Ministry of Education and Scientific Research, Institute of Atomic Physics, Ministry of Research and Innovation and Institute of Atomic Physics and Universitatea Nationala de Stiinta si Tehnologie Politehnica Bucuresti, Romania;
Ministerstvo skolstva, vyskumu, vyvoja a mladeze SR, Slovakia;
National Research Foundation of South Africa, South Africa;
Swedish Research Council (VR) and Knut \& Alice Wallenberg Foundation (KAW), Sweden;
European Organization for Nuclear Research, Switzerland;
Suranaree University of Technology (SUT), National Science and Technology Development Agency (NSTDA) and National Science, Research and Innovation Fund (NSRF via PMU-B B05F650021), Thailand;
Turkish Energy, Nuclear and Mineral Research Agency (TENMAK), Turkey;
National Academy of  Sciences of Ukraine, Ukraine;
Science and Technology Facilities Council (STFC), United Kingdom;
National Science Foundation of the United States of America (NSF) and United States Department of Energy, Office of Nuclear Physics (DOE NP), United States of America.
In addition, individual groups or members have received support from:
Czech Science Foundation (grant no. 23-07499S), Czech Republic;
FORTE project, reg.\ no.\ CZ.02.01.01/00/22\_008/0004632, Czech Republic, co-funded by the European Union, Czech Republic;
European Research Council (grant no. 950692), European Union;
Deutsche Forschungs Gemeinschaft (DFG, German Research Foundation) ``Neutrinos and Dark Matter in Astro- and Particle Physics'' (grant no. SFB 1258), Germany;
FAIR - Future Artificial Intelligence Research, funded by the NextGenerationEU program (Italy);
CONVECS project, CUP C97H23001700002 FESR 2021-2027 program, Italy.

%% file: Alice_Authorlist_2025-09-18_Opt_C.tex
\begin{flushleft} 
\small

I.J.~Abualrob\,\orcidlink{0009-0005-3519-5631}\,$^{\rm 114}$, 
S.~Acharya\,\orcidlink{0000-0002-9213-5329}\,$^{\rm 50}$, 
G.~Aglieri Rinella\,\orcidlink{0000-0002-9611-3696}\,$^{\rm 32}$, 
L.~Aglietta\,\orcidlink{0009-0003-0763-6802}\,$^{\rm 24}$, 
N.~Agrawal\,\orcidlink{0000-0003-0348-9836}\,$^{\rm 25}$, 
Z.~Ahammed\,\orcidlink{0000-0001-5241-7412}\,$^{\rm 134}$, 
S.~Ahmad\,\orcidlink{0000-0003-0497-5705}\,$^{\rm 15}$, 
I.~Ahuja\,\orcidlink{0000-0002-4417-1392}\,$^{\rm 36}$, 
Z.~Akbar$^{\rm 81}$, 
A.~Akindinov\,\orcidlink{0000-0002-7388-3022}\,$^{\rm 140}$
% ...

\section*{Affiliation Notes}

$^{\rm I}$ Deceased\\
$^{\rm II}$ Also at: Max-Planck-Institut fur Physik, Munich, Germany\\
$^{\rm III}$ Also at: Czech Technical University in Prague (CZ)\\
$^{\rm IV}$ Also at: Instituto de Fisica da Universidade de Sao Paulo\\
$^{\rm V}$ Also at: Dipartimento DET del Politecnico di Torino, Turin, Italy\\
$^{\rm VI}$ Also at: Department of Applied Physics, Aligarh Muslim University, Aligarh, India\\
$^{\rm VII}$ Also at: Institute of Theoretical Physics, University of Wroclaw, Poland\\
$^{\rm VIII}$ Also at: Facultad de Ciencias, Universidad Nacional Aut\'{o}noma de M\'{e}xico, Mexico City, Mexico\\
$^{\rm IX}$ Also at: University College of Dublin, Dublin, Ireland\\

\section*{Collaboration Institutes}

$^{1}$ A.I. Alikhanyan National Science Laboratory (Yerevan Physics Institute) Foundation, Yerevan, Armenia\\
$^{2}$ AGH University of Krakow, Cracow, Poland\\
$^{3}$ Bogolyubov Institute for Theoretical Physics, National Academy of Sciences of Ukraine, Kyiv, Ukraine\\
$^{4}$ Bose Institute, Department of Physics  and Centre for Astroparticle Physics and Space Science (CAPSS), Kolkata, India\\
$^{5}$ California Polytechnic State University, San Luis Obispo, California, United States\\
$^{6}$ Central China Normal University, Wuhan, China\\
$^{7}$ Centro de Aplicaciones Tecnol\'{o}gicas y Desarrollo Nuclear (CEADEN), Havana, Cuba\\
$^{8}$ Centro de Investigaci\'{o}n y de Estudios Avanzados (CINVESTAV), Mexico City and M\'{e}rida, Mexico\\
$^{9}$ Chicago State University, Chicago, Illinois, United States\\
$^{10}$ China Nuclear Data Center, China Institute of Atomic Energy, Beijing, China\\
$^{11}$ China University of Geosciences, Wuhan, China\\
$^{12}$ Chungbuk National University, Cheongju, Republic of Korea\\
$^{13}$ Comenius University Bratislava, Faculty of Mathematics, Physics and Informatics, Bratislava, Slovak Republic\\
$^{14}$ Creighton University, Omaha, Nebraska, United States\\
$^{15}$ Department of Physics, Aligarh Muslim University, Aligarh, India\\
$^{16}$ Department of Physics, Pusan National University, Pusan, Republic of Korea\\
$^{17}$ Department of Physics, Sejong University, Seoul, Republic of Korea\\
$^{18}$ Department of Physics, University of California, Berkeley, California, United States\\
$^{19}$ Department of Physics, University of Oslo, Oslo, Norway\\
$^{20}$ Department of Physics and Technology, University of Bergen, Bergen, Norway\\
$^{21}$ Dipartimento di Fisica, Universit\`{a} di Pavia, Pavia, Italy\\
$^{22}$ Dipartimento di Fisica dell'Universit\`{a} and Sezione INFN, Cagliari, Italy\\
$^{23}$ Dipartimento di Fisica dell'Universit\`{a} and Sezione INFN, Trieste, Italy\\
$^{24}$ Dipartimento di Fisica dell'Universit\`{a} and Sezione INFN, Turin, Italy\\
$^{25}$ Dipartimento di Fisica e Astronomia dell'Universit\`{a} and Sezione INFN, Bologna, Italy\\
$^{26}$ Dipartimento di Fisica e Astronomia dell'Universit\`{a} and Sezione INFN, Catania, Italy\\
$^{27}$ Dipartimento di Fisica e Astronomia dell'Universit\`{a} and Sezione INFN, Padova, Italy\\
$^{28}$ Dipartimento di Fisica `E.R.~Caianiello' dell'Universit\`{a} and Gruppo Collegato INFN, Salerno, Italy\\
$^{29}$ Dipartimento DISAT del Politecnico and Sezione INFN, Turin, Italy\\
$^{30}$ Dipartimento di Scienze MIFT, Universit\`{a} di Messina, Messina, Italy\\
$^{31}$ Dipartimento Interateneo di Fisica `M.~Merlin' and Sezione INFN, Bari, Italy\\
$^{32}$ European Organization for Nuclear Research (CERN), Geneva, Switzerland\\
$^{33}$ Faculty of Electrical Engineering, Mechanical Engineering and Naval Architecture, University of Split, Split, Croatia\\
$^{34}$ Faculty of Nuclear Sciences and Physical Engineering, Czech Technical University in Prague, Prague, Czech Republic\\
$^{35}$ Faculty of Physics, Sofia University, Sofia, Bulgaria\\
$^{36}$ Faculty of Science, P.J.~\v{S}af\'{a}rik University, Ko\v{s}ice, Slovak Republic\\
$^{37}$ Faculty of Technology, Environmental and Social Sciences, Bergen, Norway\\
$^{38}$ Frankfurt Institute for Advanced Studies, Johann Wolfgang Goethe-Universit\"{a}t Frankfurt, Frankfurt, Germany\\
$^{39}$ Fudan University, Shanghai, China\\
$^{40}$ Gangneung-Wonju National University, Gangneung, Republic of Korea\\
$^{41}$ Gauhati University, Department of Physics, Guwahati, India\\
$^{42}$ Helmholtz-Institut f\"{u}r Strahlen- und Kernphysik, Rheinische Friedrich-Wilhelms-Universit\"{a}t Bonn, Bonn, Germany\\
$^{43}$ Helsinki Institute of Physics (HIP), Helsinki, Finland\\
$^{44}$ High Energy Physics Group,  Universidad Aut\'{o}noma de Puebla, Puebla, Mexico\\
$^{45}$ Horia Hulubei National Institute of Physics and Nuclear Engineering, Bucharest, Romania\\
$^{46}$ HUN-REN Wigner Research Centre for Physics, Budapest, Hungary\\
$^{47}$ Indian Institute of Technology Bombay (IIT), Mumbai, India\\
$^{48}$ Indian Institute of Technology Indore, Indore, India\\
$^{49}$ INFN, Laboratori Nazionali di Frascati, Frascati, Italy\\
$^{50}$ INFN, Sezione di Bari, Bari, Italy\\
$^{51}$ INFN, Sezione di Bologna, Bologna, Italy\\
$^{52}$ INFN, Sezione di Cagliari, Cagliari, Italy\\
$^{53}$ INFN, Sezione di Catania, Catania, Italy\\
$^{54}$ INFN, Sezione di Padova, Padova, Italy\\
$^{55}$ INFN, Sezione di Pavia, Pavia, Italy\\
$^{56}$ INFN, Sezione di Torino, Turin, Italy\\
$^{57}$ INFN, Sezione di Trieste, Trieste, Italy\\
$^{58}$ Inha University, Incheon, Republic of Korea\\
$^{59}$ Institute for Gravitational and Subatomic Physics (GRASP), Utrecht University/Nikhef, Utrecht, Netherlands\\
$^{60}$ Institute of Experimental Physics, Slovak Academy of Sciences, Ko\v{s}ice, Slovak Republic\\
$^{61}$ Institute of Physics, Homi Bhabha National Institute, Bhubaneswar, India\\
$^{62}$ Institute of Physics of the Czech Academy of Sciences, Prague, Czech Republic\\
$^{63}$ Institute of Space Science (ISS), Bucharest, Romania\\
$^{64}$ Institut f\"{u}r Kernphysik, Johann Wolfgang Goethe-Universit\"{a}t Frankfurt, Frankfurt, Germany\\
$^{65}$ Instituto de Ciencias Nucleares, Universidad Nacional Aut\'{o}noma de M\'{e}xico, Mexico City, Mexico\\
$^{66}$ Instituto de F\'{i}sica, Universidade Federal do Rio Grande do Sul (UFRGS), Porto Alegre, Brazil\\
$^{67}$ Instituto de F\'{\i}sica, Universidad Nacional Aut\'{o}noma de M\'{e}xico, Mexico City, Mexico\\
$^{68}$ iThemba LABS, National Research Foundation, Somerset West, South Africa\\
$^{69}$ Jeonbuk National University, Jeonju, Republic of Korea\\
$^{70}$ Korea Institute of Science and Technology Information, Daejeon, Republic of Korea\\
$^{71}$ Laboratoire de Physique Subatomique et de Cosmologie, Universit\'{e} Grenoble-Alpes, CNRS-IN2P3, Grenoble, France\\
$^{72}$ Lawrence Berkeley National Laboratory, Berkeley, California, United States\\
$^{73}$ Lund University Department of Physics, Division of Particle Physics, Lund, Sweden\\
$^{74}$ Marietta Blau Institute, Vienna, Austria\\
$^{75}$ Nagasaki Institute of Applied Science, Nagasaki, Japan\\
$^{76}$ Nara Women{'}s University (NWU), Nara, Japan\\
$^{77}$ National and Kapodistrian University of Athens, School of Science, Department of Physics , Athens, Greece\\
$^{78}$ National Centre for Nuclear Research, Warsaw, Poland\\
$^{79}$ National Institute of Science Education and Research, Homi Bhabha National Institute, Jatni, India\\
$^{80}$ National Nuclear Research Center, Baku, Azerbaijan\\
$^{81}$ National Research and Innovation Agency - BRIN, Jakarta, Indonesia\\
$^{82}$ Niels Bohr Institute, University of Copenhagen, Copenhagen, Denmark\\
$^{83}$ Nikhef, National institute for subatomic physics, Amsterdam, Netherlands\\
$^{84}$ Nuclear Physics Group, STFC Daresbury Laboratory, Daresbury, United Kingdom\\
$^{85}$ Nuclear Physics Institute of the Czech Academy of Sciences, Husinec-\v{R}e\v{z}, Czech Republic\\
$^{86}$ Oak Ridge National Laboratory, Oak Ridge, Tennessee, United States\\
$^{87}$ Ohio State University, Columbus, Ohio, United States\\
$^{88}$ Physics department, Faculty of science, University of Zagreb, Zagreb, Croatia\\
$^{89}$ Physics Department, Panjab University, Chandigarh, India\\
$^{90}$ Physics Department, University of Jammu, Jammu, India\\
$^{91}$ Physics Program and International Institute for Sustainability with Knotted Chiral Meta Matter (WPI-SKCM$^{2}$), Hiroshima University, Hiroshima, Japan\\
$^{92}$ Physikalisches Institut, Eberhard-Karls-Universit\"{a}t T\"{u}bingen, T\"{u}bingen, Germany\\
$^{93}$ Physikalisches Institut, Ruprecht-Karls-Universit\"{a}t Heidelberg, Heidelberg, Germany\\
$^{94}$ Physik Department, Technische Universit\"{a}t M\"{u}nchen, Munich, Germany\\
$^{95}$ Politecnico di Bari and Sezione INFN, Bari, Italy\\
$^{96}$ Research Division and ExtreMe Matter Institute EMMI, GSI Helmholtzzentrum f\"ur Schwerionenforschung GmbH, Darmstadt, Germany\\
$^{97}$ Saga University, Saga, Japan\\
$^{98}$ Saha Institute of Nuclear Physics, Homi Bhabha National Institute, Kolkata, India\\
$^{99}$ School of Physics and Astronomy, University of Birmingham, Birmingham, United Kingdom\\
$^{100}$ Secci\'{o}n F\'{\i}sica, Departamento de Ciencias, Pontificia Universidad Cat\'{o}lica del Per\'{u}, Lima, Peru\\
$^{101}$ SUBATECH, IMT Atlantique, Nantes Universit\'{e}, CNRS-IN2P3, Nantes, France\\
$^{102}$ Sungkyunkwan University, Suwon City, Republic of Korea\\
$^{103}$ Suranaree University of Technology, Nakhon Ratchasima, Thailand\\
$^{104}$ Technical University of Ko\v{s}ice, Ko\v{s}ice, Slovak Republic\\
$^{105}$ The Henryk Niewodniczanski Institute of Nuclear Physics, Polish Academy of Sciences, Cracow, Poland\\
$^{106}$ The University of Texas at Austin, Austin, Texas, United States\\
$^{107}$ Universidad Aut\'{o}noma de Sinaloa, Culiac\'{a}n, Mexico\\
$^{108}$ Universidade de S\~{a}o Paulo (USP), S\~{a}o Paulo, Brazil\\
$^{109}$ Universidade Estadual de Campinas (UNICAMP), Campinas, Brazil\\
$^{110}$ Universidade Federal do ABC, Santo Andre, Brazil\\
$^{111}$ Universitatea Nationala de Stiinta si Tehnologie Politehnica Bucuresti, Bucharest, Romania\\
$^{112}$ University of Cape Town, Cape Town, South Africa\\
$^{113}$ University of Derby, Derby, United Kingdom\\
$^{114}$ University of Houston, Houston, Texas, United States\\
$^{115}$ University of Jyv\"{a}skyl\"{a}, Jyv\"{a}skyl\"{a}, Finland\\
$^{116}$ University of Kansas, Lawrence, Kansas, United States\\
$^{117}$ University of Liverpool, Liverpool, United Kingdom\\
$^{118}$ University of Science and Technology of China, Hefei, China\\
$^{119}$ University of Silesia in Katowice, Katowice, Poland\\
$^{120}$ University of South-Eastern Norway, Kongsberg, Norway\\
$^{121}$ University of Tennessee, Knoxville, Tennessee, United States\\
$^{122}$ University of the Witwatersrand, Johannesburg, South Africa\\
$^{123}$ University of Tokyo, Tokyo, Japan\\
$^{124}$ University of Tsukuba, Tsukuba, Japan\\
$^{125}$ Universit\"{a}t M\"{u}nster, Institut f\"{u}r Kernphysik, M\"{u}nster, Germany\\
$^{126}$ Universit\'{e} Clermont Auvergne, CNRS/IN2P3, LPC, Clermont-Ferrand, France\\
$^{127}$ Universit\'{e} de Lyon, CNRS/IN2P3, Institut de Physique des 2 Infinis de Lyon, Lyon, France\\
$^{128}$ Universit\'{e} de Strasbourg, CNRS, IPHC UMR 7178, F-67000 Strasbourg, France, Strasbourg, France\\
$^{129}$ Universit\'{e} Paris-Saclay, Centre d'Etudes de Saclay (CEA), IRFU, D\'{e}partment de Physique Nucl\'{e}aire (DPhN), Saclay, France\\
$^{130}$ Universit\'{e}  Paris-Saclay, CNRS/IN2P3, IJCLab, Orsay, France\\
$^{131}$ Universit\`{a} degli Studi di Foggia, Foggia, Italy\\
$^{132}$ Universit\`{a} del Piemonte Orientale, Vercelli, Italy\\
$^{133}$ Universit\`{a} di Brescia, Brescia, Italy\\
$^{134}$ Variable Energy Cyclotron Centre, Homi Bhabha National Institute, Kolkata, India\\
$^{135}$ Warsaw University of Technology, Warsaw, Poland\\
$^{136}$ Wayne State University, Detroit, Michigan, United States\\
$^{137}$ Yale University, New Haven, Connecticut, United States\\
$^{138}$ Yildiz Technical University, Istanbul, Turkey\\
$^{139}$ Yonsei University, Seoul, Republic of Korea\\
$^{140}$ Affiliated with an institute formerly covered by a cooperation agreement with CERN\\
$^{141}$ Affiliated with an international laboratory covered by a cooperation agreement with CERN.\\

\end{flushleft} 

%% file: bibliography.bib
@article{LHCb:2025msn,
    author = "Aaij, Roel and others",
    collaboration = "LHCb",
    title = "{Measurement of the $W \to \mu\nu_{\mu}$ cross-sections as a function of the muon transverse momentum in $pp$ collisions at 5.02 TeV}",
    eprint = "2509.18817",
    archivePrefix = "arXiv",
    primaryClass = "hep-ex",
    reportNumber = "LHCb-PAPER-2025-031, CERN-EP-2025-197",
    doi = "10.1007/JHEP03(2026)148",
    journal = "JHEP",
    volume = "03",
    pages = "148",
    year = "2026"
}

@article{Lonnblad:2023stc,
    author = {L\"onnblad, Leif and Shah, Harsh},
    title = "{A spatially constrained QCD colour reconnection in $\textrm{pp}$, $\textrm{p}A$, and $AA$ collisions in the Pythia8/Angantyr model}",
    eprint = "2303.11747",
    archivePrefix = "arXiv",
    primaryClass = "hep-ph",
    reportNumber = "LU-TP 23-04, MCNET-23-06",
    doi = "10.1140/epjc/s10052-023-11778-3",
    journal = "Eur. Phys. J. C",
    volume = "83",
    number = "7",
    pages = "575",
    year = "2023",
    note = "[Erratum: Eur.Phys.J.C 83, 639 (2023)]"
}

@article{ALICE-PUBLIC-2021-005,
      collaboration = "ALICE",
      title         = "{ALICE 2016-2017-2018 luminosity determination for pp
                       collisions at $\mathbf{\sqrt{{\textit s}}}$ = 13 TeV}",
      year          = "2021",
      journal = "\href{https://cds.cern.ch/record/2776672"}{ALICE-PUBLIC-2021-005}"
}

@article{ParticleDataGroup:2024cfk,
    author = "Navas, S. and others",
    collaboration = "Particle Data Group",
    title = "{Review of particle physics}",
    doi = "10.1103/PhysRevD.110.030001",
    journal = "Phys. Rev. D",
    volume = "110",
    number = "3",
    pages = "030001",
    year = "2024"
}

@article{Anastasiou:2003ds,
    author = "Anastasiou, Charalampos and Dixon, Lance J. and Melnikov, Kirill and Petriello, Frank",
    title = "{High precision QCD at hadron colliders: Electroweak gauge boson rapidity distributions at NNLO}",
    eprint = "hep-ph/0312266",
    archivePrefix = "arXiv",
    reportNumber = "SLAC-PUB-10288, UH-511-1042-03",
    doi = "10.1103/PhysRevD.69.094008",
    journal = "Phys. Rev. D",
    volume = "69",
    pages = "094008",
    year = "2004"
}

@article{Melnikov:2006kv,
    author = "Melnikov, Kirill and Petriello, Frank",
    title = "{Electroweak gauge boson production at hadron colliders through $O(\alpha_s^2)$}",
    eprint = "hep-ph/0609070",
    archivePrefix = "arXiv",
    reportNumber = "UH-511-1092-06",
    doi = "10.1103/PhysRevD.74.114017",
    journal = "Phys. Rev. D",
    volume = "74",
    pages = "114017",
    year = "2006"
}

@article{Schott:2014sea,
    author = "Schott, Matthias and Dunford, Monica",
    title = "{Review of single vector boson production in pp collisions at $\sqrt{s} = 7$ TeV}",
    eprint = "1405.1160",
    archivePrefix = "arXiv",
    primaryClass = "hep-ex",
    doi = "10.1140/epjc/s10052-014-2916-1",
    journal = "Eur. Phys. J. C",
    volume = "74",
    pages = "2916",
    year = "2014"
}

@article{Alekhin:2017kpj,
    author = {Alekhin, S. and Bl\"umlein, J. and Moch, S. and Placakyte, R.},
    title = "{Parton distribution functions, $\alpha_s$, and heavy-quark masses for LHC Run II}",
    eprint = "1701.05838",
    archivePrefix = "arXiv",
    primaryClass = "hep-ph",
    reportNumber = "DESY-16-179, DO-TH-16-13",
    doi = "10.1103/PhysRevD.96.014011",
    journal = "Phys. Rev. D",
    volume = "96",
    number = "1",
    pages = "014011",
    year = "2017"
}

@article{Hou:2019efy,
    author = "Hou, Tie-Jiun and others",
    title = "{New CTEQ global analysis of quantum chromodynamics with high-precision data from the LHC}",
    eprint = "1912.10053",
    archivePrefix = "arXiv",
    primaryClass = "hep-ph",
    reportNumber = "MSUHEP-19-025, PITT-PACC-1911, SMU-HEP-19-03",
    doi = "10.1103/PhysRevD.103.014013",
    journal = "Phys. Rev. D",
    volume = "103",
    number = "1",
    pages = "014013",
    year = "2021"
}

@article{Harland-Lang:2014zoa,
    author = "Harland-Lang, L. A. and Martin, A. D. and Motylinski, P. and Thorne, R. S.",
    title = "{Parton distributions in the LHC era: MMHT 2014 PDFs}",
    eprint = "1412.3989",
    archivePrefix = "arXiv",
    primaryClass = "hep-ph",
    reportNumber = "LCTS-2014-47, IPPP-14-97, DCPT-14-194",
    doi = "10.1140/epjc/s10052-015-3397-6",
    journal = "Eur. Phys. J. C",
    volume = "75",
    number = "5",
    pages = "204",
    year = "2015"
}

@article{NNPDF:2017mvq,
    author = "Ball, Richard D. and others",
    collaboration = "NNPDF",
    title = "{Parton distributions from high-precision collider data}",
    eprint = "1706.00428",
    archivePrefix = "arXiv",
    primaryClass = "hep-ph",
    reportNumber = "EDINBURGH-2017-08, NIKHEF-2017-006, OUTP-17-04P, TIF-UNIMI-2017-3, CAVENDISH-HEP-17-06, CERN-TH-2017-077, Edinburgh 2017/08,
  Nikhef/2017-006, OUTP-17-04P,TIF-UNIMI-2017-3",
    doi = "10.1140/epjc/s10052-017-5199-5",
    journal = "Eur. Phys. J. C",
    volume = "77",
    number = "10",
    pages = "663",
    year = "2017"
}

@article{Paukkunen:2010qg,
    author = "Paukkunen, Hannu and Salgado, Carlos A.",
    title = "{Constraints for the nuclear parton distributions from Z and W production at the LHC}",
    eprint = "1010.5392",
    archivePrefix = "arXiv",
    primaryClass = "hep-ph",
    doi = "10.1007/JHEP03(2011)071",
    journal = "JHEP",
    volume = "03",
    pages = "071",
    year = "2011"
}

@article{ALICE:2021npz,
    author = "Acharya, Shreyasi and others",
    collaboration = "ALICE",
    title = "{Observation of a multiplicity dependence in 
the $p_{\rm T}$-differential charm baryon-to-meson ratios in proton\textendash{}proton collisions at $\sqrt s$ = 13 TeV}",
    eprint = "2111.11948",
    archivePrefix = "arXiv",
    primaryClass = "nucl-ex",
    reportNumber = "CERN-EP-2021-245",
    doi = "10.1016/j.physletb.2022.137065",
    journal = "Phys. Lett. B",
    volume = "829",
    pages = "137065",
    year = "2022"
}

@article{ALICE:2020msa,
    author = "Acharya, Shreyasi and others",
    collaboration = "ALICE",
    title = "{Multiplicity dependence of J/$\psi$ production at midrapidity in pp collisions at $\sqrt{s}$ = 13 TeV}",
    eprint = "2005.11123",
    archivePrefix = "arXiv",
    primaryClass = "nucl-ex",
    reportNumber = "CERN-EP-2020-088",
    doi = "10.1016/j.physletb.2020.135758",
    journal = "Phys. Lett. B",
    volume = "810",
    pages = "135758",
    year = "2020"
}

@article{ALICE:2021zkd,
    author = "Acharya, Shreyasi and others",
    collaboration = "ALICE",
    title = "{Forward rapidity J/\ensuremath{\psi} production as a function of charged-particle multiplicity in pp collisions at $ \sqrt{s} $ = 5.02 and 13 TeV}",
    eprint = "2112.09433",
    archivePrefix = "arXiv",
    primaryClass = "nucl-ex",
    reportNumber = "CERN-EP-2021-268",
    doi = "10.1007/JHEP06(2022)015",
    journal = "JHEP",
    volume = "06",
    pages = "015",
    year = "2022"
}

@article{Weber:2018ddv,
    author = "Weber, S. G. and Dubla, A. and Andronic, A. and Morsch, A.",
    title = "{Elucidating the multiplicity dependence of $\mathrm {J}/\psi $ production in proton\textendash{}proton collisions with PYTHIA8}",
    eprint = "1811.07744",
    archivePrefix = "arXiv",
    primaryClass = "nucl-th",
    doi = "10.1140/epjc/s10052-018-6531-4",
    journal = "Eur. Phys. J. C",
    volume = "79",
    number = "1",
    pages = "36",
    year = "2019"
}

@article{Aamodt:2008zz,
    author = "Aamodt, K. and others",
    collaboration = "ALICE",
    title = "{The ALICE experiment at the CERN LHC}",
    doi = "10.1088/1748-0221/3/08/S08002",
    journal = "JINST",
    volume = "3",
    pages = "S08002",
    year = "2008"
}

@article{Abelev:2014ffa,
    author = "Abelev, Betty Bezverkhny and others",
    collaboration = "ALICE",
    title = "{Performance of the ALICE Experiment at the CERN LHC}",
    eprint = "1402.4476",
    archivePrefix = "arXiv",
    primaryClass = "nucl-ex",
    reportNumber = "CERN-PH-EP-2014-031",
    doi = "10.1142/S0217751X14300440",
    journal = "Int. J. Mod. Phys. A",
    volume = "29",
    pages = "1430044",
    year = "2014"
}

@article{Alme:2010ke,
    author = "Alme, J. and others",
    title = "{The ALICE TPC, a large 3-dimensional tracking device with fast readout for ultra-high multiplicity events}",
    eprint = "1001.1950",
    archivePrefix = "arXiv",
    primaryClass = "physics.ins-det",
    doi = "10.1016/j.nima.2010.04.042",
    journal = "Nucl. Instrum. Meth. A",
    volume = "622",
    pages = "316--367",
    year = "2010"
}

@article{Aamodt:2010aa,
    author = "Aamodt, K and others",
    collaboration = "ALICE",
    title = "{Alignment of the ALICE Inner Tracking System with cosmic-ray tracks}",
    eprint = "1001.0502",
    archivePrefix = "arXiv",
    primaryClass = "physics.ins-det",
    doi = "10.1088/1748-0221/5/03/P03003",
    journal = "JINST",
    volume = "5",
    pages = "P03003",
    year = "2010"
}

@article{ALICE:2015qqj,
    author = "Adam, Jaroslav and others",
    collaboration = "ALICE",
    title = "{Pseudorapidity and transverse-momentum distributions of charged particles in proton\textendash{}proton collisions at $\sqrt s=$ 13 TeV}",
    eprint = "1509.08734",
    archivePrefix = "arXiv",
    primaryClass = "nucl-ex",
    reportNumber = "CERN-PH-EP-2015-270, ALICE-PUBLIC-2015-005",
    doi = "10.1016/j.physletb.2015.12.030",
    journal = "Phys. Lett. B",
    volume = "753",
    pages = "319--329",
    year = "2016"
}

@article{ALICE:2012pet,
    author = "Abelev, B. and others",
    collaboration = "ALICE",
    title = "{$\rm J/\psi$ Production as a function of charged-particle multiplicity in pp collisions at $\sqrt{s} = 7$ TeV}",
    eprint = "1202.2816",
    archivePrefix = "arXiv",
    primaryClass = "hep-ex",
    doi = "10.1016/j.physletb.2012.04.052",
    journal = "Phys. Lett. B",
    volume = "712",
    pages = "165--175",
    year = "2012"
}

@article{Alioli:2008gx,
    author = "Alioli, Simone and Nason, Paolo and Oleari, Carlo and Re, Emanuele",
    title = "{NLO vector-boson production matched with shower in POWHEG}",
    eprint = "0805.4802",
    archivePrefix = "arXiv",
    primaryClass = "hep-ph",
    reportNumber = "NSF-KITP-08-74",
    doi = "10.1088/1126-6708/2008/07/060",
    journal = "JHEP",
    volume = "07",
    pages = "060",
    year = "2008"
}

@article{Dulat:2015mca,
    author = "Dulat, Sayipjamal and Hou, Tie-Jiun and Gao, Jun and Guzzi, Marco and Huston, Joey and Nadolsky, Pavel and Pumplin, Jon and Schmidt, Carl and Stump, Daniel and Yuan, C. P.",
    title = "{New parton distribution functions from a global analysis of quantum chromodynamics}",
    eprint = "1506.07443",
    archivePrefix = "arXiv",
    primaryClass = "hep-ph",
    doi = "10.1103/PhysRevD.93.033006",
    journal = "Phys. Rev. D",
    volume = "93",
    number = "3",
    pages = "033006",
    year = "2016"
}

@article{Sjostrand:2014zea,
    author = {Sj\"ostrand, Torbj\"orn and Ask, Stefan and Christiansen, Jesper R. and Corke, Richard and Desai, Nishita and Ilten, Philip and Mrenna, Stephen and Prestel, Stefan and Rasmussen, Christine O. and Skands, Peter Z.},
    title = "{An introduction to PYTHIA 8.2}",
    eprint = "1410.3012",
    archivePrefix = "arXiv",
    primaryClass = "hep-ph",
    reportNumber = "LU-TP-14-36, MCNET-14-22, CERN-PH-TH-2014-190, FERMILAB-PUB-14-316-CD, DESY-14-178, SLAC-PUB-16122",
    doi = "10.1016/j.cpc.2015.01.024",
    journal = "Comput. Phys. Commun.",
    volume = "191",
    pages = "159--177",
    year = "2015"
}

@article{OrtizVelasquez:2013ofg,
    author = "Ortiz Velasquez, Antonio and Christiansen, Peter and Cuautle Flores, Eleazar and Maldonado Cervantes, Ivonne and Pai\'c, Guy",
    title = "{Color Reconnection and Flowlike Patterns in $pp$ Collisions}",
    eprint = "1303.6326",
    archivePrefix = "arXiv",
    primaryClass = "hep-ph",
    doi = "10.1103/PhysRevLett.111.042001",
    journal = "Phys. Rev. Lett.",
    volume = "111",
    number = "4",
    pages = "042001",
    year = "2013"
}

@article{Bierlich:2014xba,
    author = {Bierlich, Christian and Gustafson, G\"osta and L\"onnblad, Leif and Tarasov, Andrey},
    title = "{Effects of Overlapping Strings in pp Collisions}",
    eprint = "1412.6259",
    archivePrefix = "arXiv",
    primaryClass = "hep-ph",
    reportNumber = "MCNET-14-27, LU-TP-14-44, JLAB-THY-15-1991",
    doi = "10.1007/JHEP03(2015)148",
    journal = "JHEP",
    volume = "03",
    pages = "148",
    year = "2015"
}

@article{Bautista:2015kwa,
    author = "Bautista, I. and T\'ellez, A. Fernandez and Ghosh, Premomoy",
    title = "{Indication of change of phase in high-multiplicity proton-proton events at LHC in String Percolation Model}",
    eprint = "1509.02278",
    archivePrefix = "arXiv",
    primaryClass = "nucl-th",
    doi = "10.1103/PhysRevD.92.071504",
    journal = "Phys. Rev. D",
    volume = "92",
    number = "7",
    pages = "071504",
    year = "2015"
}

@article{ALICE:2019dfi,
    author = "Acharya, Shreyasi and others",
    collaboration = "ALICE",
    title = "{Charged-particle production as a function of multiplicity and transverse spherocity in pp collisions at $\sqrt{s} =5.02$ and 13 TeV}",
    eprint = "1905.07208",
    archivePrefix = "arXiv",
    primaryClass = "nucl-ex",
    reportNumber = "CERN-EP-2019-094",
    doi = "10.1140/epjc/s10052-019-7350-y",
    journal = "Eur. Phys. J. C",
    volume = "79",
    number = "10",
    pages = "857",
    year = "2019"
}

@article{ALICE:2019avo,
    author = "Acharya, Shreyasi and others",
    collaboration = "ALICE",
    title = "{Multiplicity dependence of (multi-)strange hadron production in proton-proton collisions at $\sqrt{s}$ = 13 TeV}",
    eprint = "1908.01861",
    archivePrefix = "arXiv",
    primaryClass = "nucl-ex",
    reportNumber = "CERN-EP-2019-168",
    doi = "10.1140/epjc/s10052-020-7673-8",
    journal = "Eur. Phys. J. C",
    volume = "80",
    number = "2",
    pages = "167",
    year = "2020"
}

@article{ALICE:2015ikl,
    author = "Adam, Jaroslav and others",
    collaboration = "ALICE",
    title = "{Measurement of charm and beauty production at central rapidity versus charged-particle multiplicity in proton-proton collisions at $ \sqrt{s}=7 $ TeV}",
    eprint = "1505.00664",
    archivePrefix = "arXiv",
    primaryClass = "nucl-ex",
    reportNumber = "CERN-PH-EP-2015-091",
    doi = "10.1007/JHEP09(2015)148",
    journal = "JHEP",
    volume = "09",
    pages = "148",
    year = "2015"
}

@article{ALICE:2018gyx,
    author = "Acharya, Shreyasi and others",
    collaboration = "ALICE",
    title = "{Azimuthal anisotropy of heavy-flavor decay electrons in p--Pb collisions at $ \sqrt{s_{\rm NN}}$ = 5.02 TeV}",
    eprint = "1805.04367",
    archivePrefix = "arXiv",
    primaryClass = "nucl-ex",
    reportNumber = "CERN-EP-2018-119",
    doi = "10.1103/PhysRevLett.122.072301",
    journal = "Phys. Rev. Lett.",
    volume = "122",
    number = "7",
    pages = "072301",
    year = "2019"
}

@article{CMS:2018loe,
    author = "Sirunyan, A. M. and others",
    collaboration = "CMS",
    title = "{Elliptic flow of charm and strange hadrons in high-multiplicity pPb collisions at $\sqrt{s_{_\mathrm{NN}}} =$ 8.16 TeV}",
    eprint = "1804.09767",
    archivePrefix = "arXiv",
    primaryClass = "hep-ex",
    reportNumber = "CMS-HIN-17-003, CERN-EP-2018-076",
    doi = "10.1103/PhysRevLett.121.082301",
    journal = "Phys. Rev. Lett.",
    volume = "121",
    number = "8",
    pages = "082301",
    year = "2018"
}

@article{CMS:2018duw,
    author = "Sirunyan, Albert M and others",
    collaboration = "CMS",
    title = "{Observation of prompt J/$\psi$ meson elliptic flow in high-multiplicity pPb collisions at $\sqrt{s_\mathrm{NN}} =$ 8.16 TeV}",
    eprint = "1810.01473",
    archivePrefix = "arXiv",
    primaryClass = "hep-ex",
    reportNumber = "CMS-HIN-18-010, CERN-EP-2018-256",
    doi = "10.1016/j.physletb.2019.02.018",
    journal = "Phys. Lett. B",
    volume = "791",
    pages = "172--194",
    year = "2019"
}

@article{ALICE:2017smo,
    author = "Acharya, S. and others",
    collaboration = "ALICE",
    title = "{Search for collectivity with azimuthal J/$\psi$-hadron correlations in high multiplicity p--Pb collisions at $\sqrt{s_{\rm NN}}$ = 5.02 and 8.16 TeV}",
    eprint = "1709.06807",
    archivePrefix = "arXiv",
    primaryClass = "nucl-ex",
    reportNumber = "CERN-EP-2017-245",
    doi = "10.1016/j.physletb.2018.02.039",
    journal = "Phys. Lett. B",
    volume = "780",
    pages = "7--20",
    year = "2018"
}

@article{CMS:2020qul,
    author = "Sirunyan, Albert M and others",
    collaboration = "CMS",
    title = "{Studies of charm and beauty hadron long-range correlations in pp and pPb collisions at LHC energies}",
    eprint = "2009.07065",
    archivePrefix = "arXiv",
    primaryClass = "hep-ex",
    reportNumber = "CMS-HIN-19-009, CERN-EP-2020-141",
    doi = "10.1016/j.physletb.2020.136036",
    journal = "Phys. Lett. B",
    volume = "813",
    pages = "136036",
    year = "2021"
}

@article{ATLAS:2015hzw,
    author = "Aad, Georges and others",
    collaboration = "ATLAS",
    title = "{Observation of Long-Range Elliptic Azimuthal Anisotropies in $\sqrt{s}=$13 and 2.76 TeV pp collisions with the ATLAS Detector}",
    eprint = "1509.04776",
    archivePrefix = "arXiv",
    primaryClass = "hep-ex",
    reportNumber = "CERN-PH-EP-2015-251",
    doi = "10.1103/PhysRevLett.116.172301",
    journal = "Phys. Rev. Lett.",
    volume = "116",
    number = "17",
    pages = "172301",
    year = "2016"
}

@article{ATLAS:2016yzd,
    author = "Aaboud, Morad and others",
    collaboration = "ATLAS",
    title = "{Measurements of long-range azimuthal anisotropies and associated Fourier coefficients for $pp$ collisions at $\sqrt{s}=5.02$ and $13$ TeV and $p$+Pb collisions at $\sqrt{s_{\mathrm{NN}}}=5.02$ TeV with the ATLAS detector}",
    eprint = "1609.06213",
    archivePrefix = "arXiv",
    primaryClass = "nucl-ex",
    reportNumber = "CERN-EP-2016-200",
    doi = "10.1103/PhysRevC.96.024908",
    journal = "Phys. Rev. C",
    volume = "96",
    number = "2",
    pages = "024908",
    year = "2017"
}

@article{CMS:2013jlh,
    author = "Chatrchyan, Serguei and others",
    collaboration = "CMS",
    title = "{Multiplicity and Transverse Momentum Dependence of Two- and Four-Particle Correlations in pPb and PbPb Collisions}",
    eprint = "1305.0609",
    archivePrefix = "arXiv",
    primaryClass = "nucl-ex",
    reportNumber = "CMS-HIN-13-002, CERN-PH-EP-2013-077",
    doi = "10.1016/j.physletb.2013.06.028",
    journal = "Phys. Lett. B",
    volume = "724",
    pages = "213--240",
    year = "2013"
}

@article{ALICE:2013snk,
    author = "Abelev, Betty Bezverkhny and others",
    collaboration = "ALICE",
    title = "{Long-range angular correlations of $\rm \pi$, K and p in p--Pb collisions at $\sqrt{s_{\rm NN}}$ = 5.02 TeV}",
    eprint = "1307.3237",
    archivePrefix = "arXiv",
    primaryClass = "nucl-ex",
    reportNumber = "CERN-PH-EP-2013-115",
    doi = "10.1016/j.physletb.2013.08.024",
    journal = "Phys. Lett. B",
    volume = "726",
    pages = "164--177",
    year = "2013"
}

@article{CMS:2014und,
    author = "Khachatryan, Vardan and others",
    collaboration = "CMS",
    title = "{Long-range two-particle correlations of strange hadrons with charged particles in pPb and PbPb collisions at LHC energies}",
    eprint = "1409.3392",
    archivePrefix = "arXiv",
    primaryClass = "nucl-ex",
    reportNumber = "CMS-HIN-14-002, CERN-PH-EP-2014-212",
    doi = "10.1016/j.physletb.2015.01.034",
    journal = "Phys. Lett. B",
    volume = "742",
    pages = "200--224",
    year = "2015"
}

@article{PHENIX:2013ktj,
    author = "Adare, A. and others",
    collaboration = "PHENIX",
    title = "{Quadrupole Anisotropy in Dihadron Azimuthal Correlations in Central $d+$Au Collisions at $\sqrt{s_{\rm NN}}$ = 200 GeV}",
    eprint = "1303.1794",
    archivePrefix = "arXiv",
    primaryClass = "nucl-ex",
    doi = "10.1103/PhysRevLett.111.212301",
    journal = "Phys. Rev. Lett.",
    volume = "111",
    number = "21",
    pages = "212301",
    year = "2013"
}

@article{STAR:2015kak,
    author = "Adamczyk, L. and others",
    collaboration = "STAR",
    title = "{Long-range pseudorapidity dihadron correlations in $d$+Au collisions at $\sqrt{s_{\rm NN}}=200$ GeV}",
    eprint = "1502.07652",
    archivePrefix = "arXiv",
    primaryClass = "nucl-ex",
    doi = "10.1016/j.physletb.2015.05.075",
    journal = "Phys. Lett. B",
    volume = "747",
    pages = "265--271",
    year = "2015"
}

@article{PHENIX:2015idk,
    author = "Adare, A. and others",
    collaboration = "PHENIX",
    title = "{Measurements of elliptic and triangular flow in high-multiplicity $^{3}$He$+$Au collisions at $\sqrt{s_{\rm NN}} = 200$ GeV}",
    eprint = "1507.06273",
    archivePrefix = "arXiv",
    primaryClass = "nucl-ex",
    doi = "10.1103/PhysRevLett.115.142301",
    journal = "Phys. Rev. Lett.",
    volume = "115",
    number = "14",
    pages = "142301",
    year = "2015"
}

@article{ALICE:2012eyl,
    author = "Abelev, Betty and others",
    collaboration = "ALICE",
    title = "{Long-range angular correlations on the near and away side in p--Pb collisions at $\sqrt{s_{{\rm NN}}} = 5.02$ TeV}",
    eprint = "1212.2001",
    archivePrefix = "arXiv",
    primaryClass = "nucl-ex",
    reportNumber = "CERN-PH-EP-2012-359",
    doi = "10.1016/j.physletb.2013.01.012",
    journal = "Phys. Lett. B",
    volume = "719",
    pages = "29--41",
    year = "2013"
}

@article{ALICE:2023xiu,
    author = "Acharya, Shreyasi and others",
    collaboration = "ALICE",
    title = "{Inclusive and multiplicity dependent production of electrons from heavy-flavour hadron decays in pp and p--Pb collisions}",
    eprint = "2303.13349",
    archivePrefix = "arXiv",
    primaryClass = "nucl-ex",
    reportNumber = "CERN-EP-2023-041",
    doi = "10.1007/JHEP08(2023)006",
    journal = "JHEP",
    volume = "08",
    pages = "006",
    year = "2023"
}

@article{ALICE:2022qhn,
    author = "Acharya, Shreyasi and others",
    collaboration = "ALICE",
    title = "{Performance of the ALICE Electromagnetic Calorimeter}",
    eprint = "2209.04216",
    archivePrefix = "arXiv",
    primaryClass = "physics.ins-det",
    reportNumber = "CERN-EP-2022-184",
    doi = "10.1088/1748-0221/18/08/P08007",
    journal = "JINST",
    volume = "18",
    number = "08",
    pages = "P08007",
    year = "2023"
}

@article{Awes:1992yp,
    author = "Awes, T. C. and Obenshain, F. E. and Plasil, F. and Saini, S. and Sorensen, S. P. and Young, G. R.",
    title = "{A Simple method of shower localization and identification in laterally segmented calorimeters}",
    doi = "10.1016/0168-9002(92)90858-2",
    journal = "Nucl. Instrum. Meth. A",
    volume = "311",
    pages = "130--138",
    year = "1992"
}

@article{ALICE:2017nce,
    author = "Acharya, Shreyasi and others",
    collaboration = "ALICE",
    title = "{Production of ${\pi ^0}$ and $\eta $ mesons up to high transverse momentum in pp collisions at 2.76 TeV}",
    eprint = "1702.00917",
    archivePrefix = "arXiv",
    primaryClass = "hep-ex",
    reportNumber = "CERN-EP-2017-019",
    doi = "10.1140/epjc/s10052-017-4890-x",
    journal = "Eur. Phys. J. C",
    volume = "77",
    number = "5",
    pages = "339",
    year = "2017"
}

@article{ALICE:2015zhm,
    author = "Adam, Jaroslav and others",
    collaboration = "ALICE",
    title = "{Measurement of electrons from heavy-flavour hadron decays in p--Pb collisions at $\sqrt{s_{\rm NN}} =$ 5.02 TeV}",
    eprint = "1509.07491",
    archivePrefix = "arXiv",
    primaryClass = "nucl-ex",
    reportNumber = "CERN-PH-EP-2015-262",
    doi = "10.1016/j.physletb.2015.12.067",
    journal = "Phys. Lett. B",
    volume = "754",
    pages = "81--93",
    year = "2016"
}

@article{CMS:2011aa,
    author = "Chatrchyan, Serguei and others",
    collaboration = "CMS",
    title = "{Measurement of the Inclusive $W$ and $Z$ Production Cross Sections in $pp$ Collisions at $\sqrt{s}=7$ TeV}",
    eprint = "1107.4789",
    archivePrefix = "arXiv",
    primaryClass = "hep-ex",
    reportNumber = "CERN-PH-EP-2011-107, CMS-EWK-10-005",
    doi = "10.1007/JHEP10(2011)132",
    journal = "JHEP",
    volume = "10",
    pages = "132",
    year = "2011"
}

@article{CMS:2016mwa,
    author = "Khachatryan, Vardan and others",
    collaboration = "CMS",
    title = "{Measurement of the transverse momentum spectra of weak vector bosons produced in proton-proton collisions at $ \sqrt{s}=8 $ TeV}",
    eprint = "1606.05864",
    archivePrefix = "arXiv",
    primaryClass = "hep-ex",
    reportNumber = "CMS-SMP-14-012, CERN-EP-2016-152",
    doi = "10.1007/JHEP02(2017)096",
    journal = "JHEP",
    volume = "02",
    pages = "096",
    year = "2017"
}

@article{CMS:2020cph,
    author = "Sirunyan, Albert M and others",
    collaboration = "CMS",
    title = "{Measurements of the $W$ boson rapidity, helicity, double-differential cross sections, and charge asymmetry in $pp$ collisions at $\sqrt {s}$ = 13  TeV}",
    eprint = "2008.04174",
    archivePrefix = "arXiv",
    primaryClass = "hep-ex",
    reportNumber = "CMS-SMP-18-012, CERN-EP-2020-116",
    doi = "10.1103/PhysRevD.102.092012",
    journal = "Phys. Rev. D",
    volume = "102",
    number = "9",
    pages = "092012",
    year = "2020"
}

@article{ATLAS:2015iiu,
    author = "Aad, Georges and others",
    collaboration = "ATLAS",
    title = "{Measurement of the transverse momentum and $\phi ^*_{\eta }$ distributions of Drell\textendash{}Yan lepton pairs in proton\textendash{}proton collisions at $\sqrt{s}=8$  TeV with the ATLAS detector}",
    eprint = "1512.02192",
    archivePrefix = "arXiv",
    primaryClass = "hep-ex",
    reportNumber = "CERN-PH-EP-2015-275",
    doi = "10.1140/epjc/s10052-016-4070-4",
    journal = "Eur. Phys. J. C",
    volume = "76",
    number = "5",
    pages = "291",
    year = "2016"
}

@article{ATLAS:2016nqi,
    author = "Aaboud, Morad and others",
    collaboration = "ATLAS",
    title = "{Precision measurement and interpretation of inclusive $W^+$ , $W^-$ and $Z/\gamma ^*$ production cross sections with the ATLAS detector}",
    eprint = "1612.03016",
    archivePrefix = "arXiv",
    primaryClass = "hep-ex",
    reportNumber = "CERN-EP-2016-272",
    doi = "10.1140/epjc/s10052-017-4911-9",
    journal = "Eur. Phys. J. C",
    volume = "77",
    number = "6",
    pages = "367",
    year = "2017"
}

@article{ATLAS:2018pyl,
    author = "Aaboud, Morad and others",
    collaboration = "ATLAS",
    title = "{Measurements of $W$ and $Z$ boson production in $pp$ collisions at $\sqrt{s}=5.02$ TeV with the ATLAS detector}",
    eprint = "1810.08424",
    archivePrefix = "arXiv",
    primaryClass = "hep-ex",
    reportNumber = "CERN-EP-2018-259",
    doi = "10.1140/epjc/s10052-019-6622-x",
    journal = "Eur. Phys. J. C",
    volume = "79",
    number = "2",
    pages = "128",
    year = "2019",
    note = "[Erratum: Eur.Phys.J.C 79, 374 (2019)]"
}

@article{ATLAS:2019fyu,
    author = "Aad, Georges and others",
    collaboration = "ATLAS",
    title = "{Measurement of $W^{\pm }$-boson and Z-boson production cross-sections in pp collisions at $\sqrt{s}=2.76$ TeV with the ATLAS detector}",
    eprint = "1907.03567",
    archivePrefix = "arXiv",
    primaryClass = "hep-ex",
    reportNumber = "CERN-EP-2019-095",
    doi = "10.1140/epjc/s10052-019-7399-7",
    journal = "Eur. Phys. J. C",
    volume = "79",
    number = "11",
    pages = "901",
    year = "2019"
}

@article{Yan:2022pzl,
    author = "Yan, Mengshi and Hou, Tie-Jiun and Nadolsky, Pavel and Yuan, C. -P.",
    title = "{CT18 global PDF fit at leading order in QCD}",
    eprint = "2205.00137",
    archivePrefix = "arXiv",
    primaryClass = "hep-ph",
    reportNumber = "MSUHEP-22-013, SMU-HEP-22-05",
    doi = "10.1103/PhysRevD.107.116001",
    journal = "Phys. Rev. D",
    volume = "107",
    number = "11",
    pages = "116001",
    year = "2023"
}

@article{NNPDF:2021njg,
    author = "Ball, Richard D. and others",
    collaboration = "NNPDF",
    title = "{The path to proton structure at 1\% accuracy}",
    eprint = "2109.02653",
    archivePrefix = "arXiv",
    primaryClass = "hep-ph",
    reportNumber = "Edinburgh 2021/12, Nikhef-2021-013, TIF-UNIMI-2021-11",
    doi = "10.1140/epjc/s10052-022-10328-7",
    journal = "Eur. Phys. J. C",
    volume = "82",
    number = "5",
    pages = "428",
    year = "2022"
}

@article{Sjostrand:2004pf,
    author = "Sjostrand, T. and Skands, Peter Z.",
    title = "{Multiple interactions and the structure of beam remnants}",
    eprint = "hep-ph/0402078",
    archivePrefix = "arXiv",
    reportNumber = "LU-TP-04-08",
    doi = "10.1088/1126-6708/2004/03/053",
    journal = "JHEP",
    volume = "03",
    pages = "053",
    year = "2004"
}

@article{Gustafson:2009qz,
    author = "Gustafson, Gosta",
    editor = "Golec-Biernat, Krzysztof and Praszalowicz, Michal",
    title = "{Multiple Interactions, Saturation, and Final States in pp Collisions and DIS}",
    eprint = "0905.2492",
    archivePrefix = "arXiv",
    primaryClass = "hep-ph",
    journal = "Acta Phys. Polon. B",
    volume = "40",
    pages = "1981--1996",
    year = "2009"
}

@article{Sarma:2021qfr,
    author = "Sarma, Pranjal and Barman, Banajit and Bhattacharjee, Buddhadeb",
    title = "{Effect of color reconnection on multiplicity dependent charged particle production in PYTHIA-generated pp collisions at the LHC energies}",
    eprint = "2107.13732",
    archivePrefix = "arXiv",
    primaryClass = "hep-ph",
    doi = "10.1140/epja/s10050-023-00989-7",
    journal = "Eur. Phys. J. A",
    volume = "59",
    number = "4",
    pages = "76",
    year = "2023"
}

@article{ATLAS:2017wln,
    collaboration = "ATLAS",
    title = "{A study of different colour reconnection settings for Pythia8 generator using underlying event observables}",
    journal  = "ATL-PHYS-PUB-2017-008",
    url           = "https://cds.cern.ch/record/2262253",
    month = "5",
    year = "2017"
}

@book{Brun:1082634,
      author        = "Brun, René and Bruyant, F and Carminati, Federico and
                       Giani, Simone and Maire, M and McPherson, A and Patrick, G
                       and Urban, L",
      title         = "{GEANT: Detector Description and Simulation Tool; Oct
                       1994}",
      publisher     = "CERN",
      address       = "Geneva",
      series        = "CERN Program Library",
      year          = "1993",
      url           = "http://cds.cern.ch/record/1082634",
      note          = "Long Writeup W5013",
      doi           = "10.17181/CERN.MUHF.DMJ1",
}

@article{ALICE:2013axi,
    author = "Abbas, E. and others",
    collaboration = "ALICE",
    title = "{Performance of the ALICE VZERO system}",
    eprint = "1306.3130",
    archivePrefix = "arXiv",
    primaryClass = "nucl-ex",
    reportNumber = "CERN-PH-EP-2013-082",
    doi = "10.1088/1748-0221/8/10/P10016",
    journal = "JINST",
    volume = "8",
    pages = "P10016",
    year = "2013"
}

@article{LHCb:2014liz,
    author = "Aaij, Roel and others",
    collaboration = "LHCb",
    title = "{Measurement of the forward $W$ boson cross-section in $pp$ collisions at $\sqrt{s} = 7 {\rm \, TeV}$}",
    eprint = "1408.4354",
    archivePrefix = "arXiv",
    primaryClass = "hep-ex",
    reportNumber = "LHCB-PAPER-2014-033, CERN-PH-EP-2014-175",
    doi = "10.1007/JHEP12(2014)079",
    journal = "JHEP",
    volume = "12",
    pages = "079",
    year = "2014"
}

@article{LHCb:2015mad,
    author = "Aaij, Roel and others",
    collaboration = "LHCb",
    title = "{Measurement of forward W and Z boson production in $pp$ collisions at $ \sqrt{s}=8 $ TeV}",
    eprint = "1511.08039",
    archivePrefix = "arXiv",
    primaryClass = "hep-ex",
    reportNumber = "LHCB-PAPER-2015-049, CERN-PH-EP-2015-301",
    doi = "10.1007/JHEP01(2016)155",
    journal = "JHEP",
    volume = "01",
    pages = "155",
    year = "2016"
}

@article{LHCb:2016fbk,
    author = "Aaij, Roel and others",
    collaboration = "LHCb",
    title = "{Measurement of the forward Z boson production cross-section in pp collisions at $\sqrt{s} = 13$ TeV}",
    eprint = "1607.06495",
    archivePrefix = "arXiv",
    primaryClass = "hep-ex",
    reportNumber = "LHCB-PAPER-2016-021, CERN-EP-2016-170",
    doi = "10.1007/JHEP09(2016)136",
    journal = "JHEP",
    volume = "09",
    pages = "136",
    year = "2016"
}

@article{LHCb:2015okr,
    author = "Aaij, Roel and others",
    collaboration = "LHCb",
    title = "{Measurement of the forward $Z$ boson production cross-section in $pp$ collisions at $\sqrt{s}=7$ TeV}",
    eprint = "1505.07024",
    archivePrefix = "arXiv",
    primaryClass = "hep-ex",
    reportNumber = "LHCB-PAPER-2015-001, CERN-PH-EP-2015-102",
    doi = "10.1007/JHEP08(2015)039",
    journal = "JHEP",
    volume = "08",
    pages = "039",
    year = "2015"
}

@article{LHCb:2021huf,
    author = "Aaij, R. and others",
    collaboration = "LHCb",
    title = "{Precision measurement of forward $Z$ boson production in proton-proton collisions at $\sqrt{s} = 13$ TeV}",
    eprint = "2112.07458",
    archivePrefix = "arXiv",
    primaryClass = "hep-ex",
    reportNumber = "LHCb-PAPER-2021-037, CERN-EP-2021-246",
    doi = "10.1007/JHEP07(2022)026",
    journal = "JHEP",
    volume = "07",
    pages = "026",
    year = "2022"
}

@article{LHCb:2023qav,
    author = "Aaij, Roel and others",
    collaboration = "LHCb",
    title = "{Measurement of the Z boson production cross-section in pp collisions at $ \sqrt{s} $ = 5.02 TeV}",
    eprint = "2308.12940",
    archivePrefix = "arXiv",
    primaryClass = "hep-ex",
    reportNumber = "LHCb-PAPER-2023-010, CERN-EP-2023-141",
    doi = "10.1007/JHEP02(2024)070",
    journal = "JHEP",
    volume = "02",
    pages = "070",
    year = "2024"
}

@article{Bazavov:2011nk,
    author = "Bazavov, A. and others",
    title = "{The chiral and deconfinement aspects of the QCD transition}",
    eprint = "1111.1710",
    archivePrefix = "arXiv",
    primaryClass = "hep-lat",
    doi = "10.1103/PhysRevD.85.054503",
    journal = "Phys. Rev. D",
    volume = "85",
    pages = "054503",
    year = "2012"
}

@article{PHENIX:2023dxl,
    author = "Abdulameer, N. J. and others",
    collaboration = "PHENIX",
    title = "{Disentangling centrality bias and final-state effects in the production of high-$p_T$$\pi^0$ using direct $\gamma$ in $d$$+$Au collisions at $\sqrt{s_{\rm NN}}$ = 200 GeV}",
    eprint = "2303.12899",
    archivePrefix = "arXiv",
    primaryClass = "nucl-ex",
    doi = "10.1103/PhysRevLett.134.022302",
    journal = "Phys. Rev. Lett.",
    volume = "134",
    number = "2",
    pages = "022302",
    year = "2025"
}

@article{Sjostrand:2006za,
    author = "Sjostrand, Torbjorn and Mrenna, Stephen and Skands, Peter Z.",
    title = "{PYTHIA 6.4 Physics and Manual}",
    eprint = "hep-ph/0603175",
    archivePrefix = "arXiv",
    reportNumber = "FERMILAB-PUB-06-052-CD-T, LU-TP-06-13",
    doi = "10.1088/1126-6708/2006/05/026",
    journal = "JHEP",
    volume = "05",
    pages = "026",
    year = "2006"
}

@article{ATLAS:2016fij,
    author = "Aad, Georges and others",
    collaboration = "ATLAS",
    title = "{Measurement of $W^{\pm}$ and $Z$-boson production cross sections in $pp$ collisions at $\sqrt{s}=13$ TeV with the ATLAS detector}",
    eprint = "1603.09222",
    archivePrefix = "arXiv",
    primaryClass = "hep-ex",
    reportNumber = "CERN-EP-2016-069",
    doi = "10.1016/j.physletb.2016.06.023",
    journal = "Phys. Lett. B",
    volume = "759",
    pages = "601--621",
    year = "2016"
}

@article{CMS:2024gzs,
    author = "Hayrapetyan, Aram and others",
    collaboration = "CMS",
    title = "{Stairway to discovery: a report on the CMS programme of cross section measurements from millibarns to femtobarns}",
    eprint = "2405.18661",
    archivePrefix = "arXiv",
    primaryClass = "hep-ex",
    reportNumber = "CMS-SMP-23-004, CERN-EP-2024-117",
    doi = "10.1016/j.physrep.2024.11.005",
    journal = "Phys. Rept.",
    volume = "1115",
    pages = "3--115",
    year = "2025"
}

@article{ALICE:2020swj,
    author = "Acharya, Shreyasi and others",
    collaboration = "ALICE",
    title = "{Pseudorapidity distributions of charged particles as a function of mid- and forward rapidity multiplicities in pp collisions at $\sqrt{s}$~=~5.02, 7 and 13 TeV}",
    eprint = "2009.09434",
    archivePrefix = "arXiv",
    primaryClass = "nucl-ex",
    reportNumber = "CERN-EP-2020-170",
    doi = "10.1140/epjc/s10052-021-09349-5",
    journal = "Eur. Phys. J. C",
    volume = "81",
    number = "7",
    pages = "630",
    year = "2021"
}

@article{ALICE:2022cxs,
    author = "Acharya, Shreyasi and others",
    collaboration = "ALICE",
    title = "{W$^\pm$-boson production in p$-$Pb collisions at $\sqrt{s_{NN}} = 8.16$ TeV and PbPb collisions at $\sqrt{s_{NN}} = 5.02$ TeV}",
    eprint = "2204.10640",
    archivePrefix = "arXiv",
    primaryClass = "nucl-ex",
    reportNumber = "CERN-EP-2022-076",
    doi = "10.1007/JHEP05(2023)036",
    journal = "JHEP",
    volume = "05",
    pages = "036",
    year = "2023"
}
